\documentclass[5p]{elsarticle}
\makeatletter
\def\ps@pprintTitle{%
 \let\@oddhead\@empty
 \let\@evenhead\@empty
 \def\@oddfoot{}%
 \let\@evenfoot\@oddfoot}
\makeatother

\usepackage{csquotes}
\usepackage{graphicx,color}
\usepackage{float}
\usepackage[caption=false]{subfig}
\usepackage{amsmath}
\usepackage{amssymb}
\usepackage{multirow}
\usepackage{dsfont}
\usepackage{adjustbox}
\usepackage{pdflscape}

\begin{document}

\title{Retarded room temperature Hamaker coefficients between bulk elemental metals\vspace*{-3.32mm}}
\author{P. Tolias}
\address{Space and Plasma Physics - KTH Royal Institute of Technology, Teknikringen 31, 10044 Stockholm, Sweden\vspace*{-8.32mm}}
\begin{abstract}
\noindent The exact Lifshitz theory is utilized for the systematic calculation of the retarded room temperature Hamaker coefficients between $26$ identical isotropic elemental metals embedded in vacuum or pure water. The full spectral method, complemented with a Drude model low frequency extrapolation, is employed for metals with input from extended-in-frequency dielectric data, while the simple spectral method is employed for water with input from three dielectric representations. The accuracy of common Lifshitz approximations is quantified. A novel compact semi-empirical expression is proposed for the separation-dependence of the Hamaker coefficient that is demonstrated to be very accurate for all combinations.
\end{abstract}
\maketitle

\section{Introduction}\label{intro}

\noindent The van der Waals force and its omnipresent dispersion force component is a ubiquitous interaction between atoms, molecules and condensed matter objects that is of electromagnetic origin and does not involve the overlapping of electronic wavefunctions (see chemical bonding)\,\cite{surface1,Lifshit6,Lifshit7}. Such weak long range surface forces are important for bodies that lie in close proximity (of the order ten to hundred nanometers) and play a central role in numerous colloidal and interfacial phenomena. Early theoretical progress in van der Waals interactions dates back to the London theory of the dispersion component\,\cite{LondonDi}, the Hamaker pairwise summation method\,\cite{HamakerP} and the Casimir-Polder analysis of the asymptotic limit\,\cite{CasimirP}, while the advent of the celebrated Lifshitz theory unified and generalized their description\,\cite{Lifshit1}. Finally, it is worth pointing out that macroscopic quantum electrodynamics has emerged as the most powerful formalism for the calculation of dispersion forces being applicable to realistic materials, non-trivial environments, complex geometries and even non-equilibrium systems\,\cite{Buhmann1,Buhmann2,Buhmann3}.

Lifshitz theory allows for a rigorous calculation of dispersion forces\,\cite{Lifshit1,Lifshit2,Lifshit3,Lifshit4,Lifshit5,Lifchem1,Lifchem2,Lifquan1,Lifquan2,Lifquan3} that includes both thermal and quantum fluctuations of the electromagnetic field. In the Lifshitz formalism, the Hamaker coefficient or Hamaker function emerges isolating all complexities which originate from cumulative interactions between the instantaneously induced or the permanent multipoles arising inside the bodies and mediated by the ambient medium, including the retardation of the electromagnetic interaction. Ultimately, the Hamaker coefficient depends on the system temperature (assumption of thermodynamic equilibrium), on the body separation and on the relative spectral mismatches in the magneto-dielectric responses between the bodies and the intervening medium. The latter dependence is rather obscured by relativistic effects, but becomes evident in the non-retarded limit where the Hamaker coefficient collapses to a Hamaker constant.

Most available Lifshitz theory calculations focus on the non-retarded zero separation limit\,\cite{Lifshit6,Lifshit7,LifNret1,LifNret2,LifNret3,LifNret4,LifNret5,LifNret6,intromin}, which is the most important for colloidal stability\,\cite{introd01,introd02}, for powder adhesion\,\cite{introd03,introd04,introd05}, for liquid solid wettability\,\cite{introd06,introd07} as well as for thin liquid film stability \& evolution\,\cite{introd08,introd09}. On the other hand, the fully retarded long separation limit, whose non-entropic component represents the Casimir-Polder force, has also gathered lots of attention mainly due to its connection with the zero-point fluctuations in vacuum\,\cite{Lifquan2,Lifquan3,introd10,introd11}. Unfortunately, the intermediate separation limit where relativistic effects are important but not dominant has been much less explored. Despite the importance of this regime for numerous physical scenarios and technological applications such as biological interactions\,\cite{introd12}, gas hydrate self-preservation through ice layer formation\,\cite{introd13,introd14} and nanosystem design or manipulation\,\cite{introd15}, there exist few dedicated computational investigations available in the literature\,\cite{LifretP1,LifretP2,LifretP3,LifretP4}.

The present contribution targets the small and intermediate separation limit. Given technological and medical applications of metallic nano- and micro-particles, the interacting objects are different elemental metals embedded in vacuum or pure water. In particular, exact Lifshitz calculations are reported for the retarded room temperature Hamaker coefficients between $26$ identical isotropic polycrystalline metals in vacuum or pure water for separations mainly within $[0,200]\,$nm but also within $[0,1000]\,$nm. The necessary imaginary argument dielectric function is computed with the full spectral method combined with a Drude model low frequency extrapolation for metals and with the simple spectral method combined with established dielectric parameterizations for pure water. Retarded room temperature Hamaker coefficients are also computed with various common approximations of Lifshitz theory aiming to establish their level of accuracy. A compact semi-empirical expression is also constructed that features only four least-square fitted parameters and describes the separation dependence of the retarded room temperature Hamaker coefficient between metals embedded in vacuum or pure water.

\section{Theoretical aspects}\label{theoretical}

\subsection{Exact retarded Hamaker coefficients}\label{LifshitzTheoryGeneral}

\noindent Let us consider the interactions between two homogeneous isotropic infinitely thick infinitely extended parallel (semi-spaces) solid bodies (labelled by the indices $1,2$) that are separated by a homogeneous isotropic medium (labelled by index $3$). The so-called retarded Lifshitz theory yields the Hamaker form $g_{132}(l)=-A_{132}(l)/(12\pi{l}^2)$\,\cite{HamakerP,Lifshit1,Lifshit2,Lifshit3,Lifshit4} for the van der Waals interaction free energy per unit area, where $l$ is the separation and $A_{132}(l)$ is the Hamaker coefficient, whose most general representation reads as\,\cite{Lifshit5}
\begin{align*}
&A_{132}(l)=-\frac{3}{2}k_{\mathrm{b}}T{\sum_{n=0}^{\infty}}^{\prime}\int_{r(l,\xi_n)}^{\infty}x\ln\left\{\left[1-\bar{\Delta}_{13}(l,x,\xi_n)\times\right.\right.\\&\,\,\,\left.\left.\bar{\Delta}_{23}(l,x,\xi_n)e^{-x}\right]\left[1-\Delta_{13}(l,x,\xi_n)\Delta_{23}(l,x,\xi_n)e^{-x}\right]\right\}dx\,,
\end{align*}
where the auxiliary functions $\bar{\Delta}_{ij}(\cdot),\,\Delta_{ij}(\cdot),\,x_i(\cdot),\,r_n(\cdot)$ are defined by
\begin{align*}
&\bar{\Delta}_{ij}(l,x,\xi_n)=\frac{x_j(l,x,\xi_n)\epsilon_i(\imath\xi_n)-x_i(l,x,\xi_n)\epsilon_j(\imath\xi_n)}{x_j(l,x,\xi_n)\epsilon_i(\imath\xi_n)+x_i(l,x,\xi_n)\epsilon_j(\imath\xi_n)}\,,\\
&\Delta_{ij}(l,x,\xi_n)=\frac{x_j(l,x,\xi_n)\mu_i(\imath\xi_n)-x_i(l,x,\xi_n)\mu_j(\imath\xi_n)}{x_j(l,x,\xi_n)\mu_i(\imath\xi_n)+x_i(l,x,\xi_n)\mu_j(\imath\xi_n)}\,,\\
&x_i^2(l,x,\xi_n)=x^2+\left(\frac{2l\xi_n}{c}\right)^2\left[\epsilon_i(\imath\xi_n)\mu_i(\imath\xi_n)-\epsilon_3(\imath\xi_n)\mu_3(\imath\xi_n)\right]\,,\\
&r(l,\xi_n)=2l\frac{\sqrt{\epsilon_3(\imath\xi_n)\mu_3(\imath\xi_n)}}{c}\xi_n\,.
\end{align*}
In the above; $\epsilon_j(\imath\xi_n)$ and $\mu_j(\imath\xi_n)$ denote the relative dielectric permittivity and magnetic permeability of imaginary argument evaluated at the bosonic Matsubara frequencies $\xi_n=2\pi{n}k_{\mathrm{b}}T/\hbar$, the prime above the series indicates that $n=0$ is considered with a half-weight, $T$ the temperature, $k_{\mathrm{b}}$ the Boltzmann constant, $c$ the speed of light in vacuum.

Let us now confine the discussion to \emph{conducting bodies that are surrounded by a non-conducting medium}. General properties of the dielectric and magnetic responses allow some simplifications in the static $n=0$ and dynamic $n\neq0$ contributions. Courtesy of the logarithm product rule, the static term can be split into a dielectric and a magnetic contribution. Moreover, $\bar{\Delta}_{ij}(l,x,\xi_n=0)=1$ in the static dielectric contribution since $\epsilon_1(0),\epsilon_2(0)\gg\epsilon_3(0)$, whereas $\Delta_{ij}(l,x,\xi_n=0)=\Delta_{ij}^{\mathrm{s}}$ in the static magnetic contribution with $\Delta_{ij}^{\mathrm{s}}$ easy to express via volume magnetic susceptibilities. In addition, the logarithmic factor of each contribution is expanded in a Taylor series, the series operators are interchanged with the integral operators and the emerging integral is computed analytically. The series $\sum_{m=1}^{\infty}m^{-3}=\zeta(3)$ and  $\sum_{m=1}^{\infty}(\Delta_{13}^{\mathrm{s}}\Delta_{23}^{\mathrm{s}})^mm^{-3}=\mathrm{Li}_3(\Delta_{13}^{\mathrm{s}}\Delta_{23}^{\mathrm{s}})$ emerge for the static dielectric and magnetic contributions, respectively, with $\zeta(\cdot)$ the Riemann $\zeta$-function and with $\mathrm{Li}_3(\cdot)$ the polylogarithm function of the third order\,\cite{magneti1}. Finally, the magnetic permeability is typically characterized by a single relaxation frequency, which lies well below the first room temperature Matsubara frequency\,\cite{magneti2,magneti3}, yielding $\mu_i(\imath\xi_n\neq0)=1$. Overall, the Hamaker coefficient becomes
\begin{align}
&A_{132}(l)=-\frac{3}{2}k_{\mathrm{b}}T{\sum_{n=1}^{\infty}}\int_{r(l,\xi_n)}^{\infty}x\ln\left\{\left[1-\bar{\Delta}_{13}(l,x,\xi_n)\times\right.\right.\nonumber\\&\,\,\,\left.\left.\bar{\Delta}_{23}(l,x,\xi_n)e^{-x}\right]\left[1-\Delta_{13}(l,x,\xi_n)\Delta_{23}(l,x,\xi_n)e^{-x}\right]\right\}dx\nonumber\\&\,\,\,+\frac{3}{4}k_{\mathrm{b}}T\zeta(3)+\frac{3}{4}k_{\mathrm{b}}T\mathrm{Li}_3(\Delta_{13}^{\mathrm{s}}\Delta_{23}^{\mathrm{s}})\label{Hamakergeneral1},
\end{align}
where $\Delta_{ij}^{\mathrm{s}},\,\bar{\Delta}_{ij}(\cdot),\,\Delta_{ij}(\cdot),\,x_i(\cdot),\,r_n(\cdot)$ are now defined by
\begin{align}
&\Delta_{ij}^{\mathrm{s}}=\Delta_{ij}(l,x,\xi_n=0)=\frac{\chi_{\mathrm{v},i}-\chi_{\mathrm{v},j}}{(2\pi)^{-1}+\chi_{\mathrm{v},i}+\chi_{\mathrm{v},j}},\label{Hamakergeneral2}\\
&\bar{\Delta}_{ij}(l,x,\xi_n)=\frac{x_j(l,x,\xi_n)\epsilon_i(\imath\xi_n)-x_i(l,x,\xi_n)\epsilon_j(\imath\xi_n)}{x_j(l,x,\xi_n)\epsilon_i(\imath\xi_n)+x_i(l,x,\xi_n)\epsilon_j(\imath\xi_n)},\label{Hamakergeneral3}\\
&\Delta_{ij}(l,x,\xi_n)=\frac{x_j(l,x,\xi_n)-x_i(l,x,\xi_n)}{x_j(l,x,\xi_n)+x_i(l,x,\xi_n)}\label{Hamakergeneral4},\\
&x_i(l,x,\xi_n)=\sqrt{x^2+\left(\frac{2l\xi_n}{c}\right)^2\left[\epsilon_i(\imath\xi_n)-\epsilon_3(\imath\xi_n)\right]},\label{Hamakergeneral5}\\
&r(l,\xi_n)=2l\frac{\sqrt{\epsilon_3(\imath\xi_n)}}{c}\xi_n\label{Hamakergeneral6},
\end{align}
where $\chi_{\mathrm{v}}$ denotes the volume magnetic susceptibility that is connected with the relative static magnetic permeability through $\mu=1+4\pi\chi_{\mathrm{v}}$\,\cite{magneti4}.

\subsection{Approximate retarded Hamaker coefficients}\label{LifshitzTheoryApproximate}

\noindent Aiming to reduce the high computational cost that originates mainly from the infinite Matsubara series but also from the complicated logarithmic integrand, different approximate forms of the retarded Hamaker coefficient have been discussed in the literature\,\cite{Lifshit5}. It is instructive to derive the most common Lifshitz theory approximations and to briefly discuss their physical meaning. It is pointed out that these approximations are more computationally cheap in the non-retarded limit, since the inclusion of retardation effects necessarily increases the numerical complexity.

Within the \emph{low temperature approximation}\,\cite{Lifshit6,Lifshit2}, it is implicitly assumed that the instantaneously induced multipoles are predominantly of quantum nature and not of thermal nature. In other words, the spacing between successive bosonic Matsubara frequencies is considered to be infinitesimally small, which formally allows for the replacement of the discrete spectrum $\xi_n$ with a continuous spectrum $\xi$. The starting point is the general retarded Hamaker form with the static dielectric and magnetic contributions isolated, see Eq.(\ref{Hamakergeneral1}). Then, the series operator is replaced with an integral operator, the series index is transformed to the integration interval $(2\pi{k}_{\mathrm{b}}T/\hbar,\infty)$ and the substitution $1\to{d}n=[\hbar/(2\pi{k}_{\mathrm{b}}T)]d\xi$ is employed. Overall,
\begin{align}
&A_{132}^{\mathrm{lt}}(l)=-\frac{3\hbar}{4\pi}\int_{\frac{2\pi{k}_{\mathrm{b}}T}{\hbar}}^{\infty}\int_{r(l,\xi)}^{\infty}x\ln\left\{\left[1-\bar{\Delta}_{13}(l,x,\xi)\right.\right.\nonumber\\&\,\,\,\left.\left.\times\bar{\Delta}_{23}(l,x,\xi)e^{-x}\right]\left[1-\Delta_{13}(l,x,\xi)\Delta_{23}(l,x,\xi)e^{-x}\right]\right\}dxd\xi\nonumber\\&\,\,\,\,+\frac{3}{4}k_{\mathrm{b}}T\zeta(3)+\frac{3}{4}k_{\mathrm{b}}T\mathrm{Li}_3(\Delta_{13}^{\mathrm{s}}\Delta_{23}^{\mathrm{s}})\label{Hamakerlowtemp1},
\end{align}
where $\Delta_{ij}^{\mathrm{s}}$ is given by Eq.(\ref{Hamakergeneral2}) and where the auxiliary functions $\bar{\Delta}_{ij}(\cdot)$, $\Delta_{ij}(\cdot)$, $x_i(\cdot)$, $r(\cdot)$ are now defined by
\begin{align}
&\bar{\Delta}_{ij}(l,x,\xi)=\frac{x_j(l,x,\xi)\epsilon_i(\imath\xi)-x_i(l,x,\xi)\epsilon_j(\imath\xi)}{x_j(l,x,\xi)\epsilon_i(\imath\xi)+x_i(l,x,\xi)\epsilon_j(\imath\xi)},\label{Hamakerlowtemp2}\\
&\Delta_{ij}(l,x,\xi)=\frac{x_j(l,x,\xi)-x_i(l,x,\xi)}{x_j(l,x,\xi)+x_i(l,x,\xi)}\label{Hamakerlowtemp3},\\
&x_i(l,x,\xi)=\sqrt{x^2+\left(\frac{2l\xi}{c}\right)^2\left[\epsilon_i(\imath\xi)-\epsilon_3(\imath\xi)\right]},\label{Hamakerlowtemp4}\\
&r(l,\xi)=2l\frac{\sqrt{\epsilon_3(\imath\xi)}}{c}\xi\label{Hamakerlowtemp5}.
\end{align}

Within the \emph{modified low temperature approximation}\,\cite{Lifshit2,Lifshit3}, the discrete frequency spectrum is still replaced with a continuous frequency spectrum. However, the replacement is applied from the dc frequency rather than from the first bosonic Matsubara frequency. As a consequence, the static dielectric contribution is incorporated in the double integral (in order to avoid double counting) and only the static magnetic contribution is isolated. This leads to the Hamaker coefficient
\begin{align}
&A_{132}^{\mathrm{lt,mod}}(l)=-\frac{3\hbar}{4\pi}\int_{0}^{\infty}\int_{r(l,\xi)}^{\infty}x\ln\left\{\left[1-\bar{\Delta}_{13}(l,x,\xi)\right.\right.\nonumber\\&\,\,\,\left.\left.\times\bar{\Delta}_{23}(l,x,\xi)e^{-x}\right]\left[1-\Delta_{13}(l,x,\xi)\Delta_{23}(l,x,\xi)e^{-x}\right]\right\}dxd\xi\nonumber\\&\,\,\,\,+\frac{3}{4}k_{\mathrm{b}}T\mathrm{Li}_3(\Delta_{13}^{\mathrm{s}}\Delta_{23}^{\mathrm{s}})\label{Hamakerlowtempmod1},
\end{align}
where $\Delta_{ij}^{\mathrm{s}}$, $\bar{\Delta}_{ij}(\cdot)$, $\Delta_{ij}(\cdot)$, $x_i(\cdot)$ and $r(\cdot)$ are still defined by Eqs.(\ref{Hamakergeneral2},\ref{Hamakerlowtemp2},\ref{Hamakerlowtemp3},\ref{Hamakerlowtemp4},\ref{Hamakerlowtemp5}), respectively.

Within the \emph{dipole approximation}\,\cite{Lifshit6,Lifshit7}, it is implicitly assumed that the instantaneously induced dipoles are dominant over instantaneously induced higher-order multipoles. The formal starting point is the general retarded Hamaker form with the static dielectric and magnetic contributions isolated, see Eq.(\ref{Hamakergeneral1}). Then, the integration variable change $p=x/r(l,\xi_n)$ is introduced, which is always permissible since $r(l,\xi_n\neq0)>0$ and removes the separation dependence from the auxiliary functions $\bar{\Delta}_{ij}$ and $\Delta_{ij}$. Afterwards, the integral is split into two integrals courtesy of the logarithm product rule and each logarithmic factor is expanded in a Taylor series $\ln{(1-y)}=-\sum_{m=1}^{\infty}y^m/m$. Finally, only the first order $m=1$ term is considered in each Taylor expansion and the two integrals are unified to a single integral. One ends up with a Hamaker coefficient of the form
\begin{align}
A_{132}^{\mathrm{dp}}(l)&=+\frac{3}{2}k_{\mathrm{b}}T\sum_{n=1}^{\infty}r^2_n(l,\xi_n)\int_{1}^{\infty}p\left[\bar{\Delta}_{13}(p,\xi_n)\bar{\Delta}_{23}(p,\xi_n)\right.\nonumber\\&\,\,\,\,\,\,\,\left.+\Delta_{13}(p,\xi_n)\Delta_{23}(p,\xi_n)\right]e^{-r(l,\xi_n)p}dp+\frac{3}{4}k_{\mathrm{b}}T\zeta(3)\nonumber\\&\,\,\,\,\,\,\,+\frac{3}{4}k_{\mathrm{b}}T\mathrm{Li}_3\left(\Delta_{13}^{\mathrm{s}}\Delta_{23}^{\mathrm{s}}\right)\label{Hamakerdipole1}\,,
\end{align}
where $\Delta_{ij}^{\mathrm{s}}$, $r(\cdot)$ are given by Eqs.(\ref{Hamakergeneral2},\ref{Hamakergeneral6}), respectively, while the auxiliary functions $\bar{\Delta}_{ij}(\cdot)$, $\Delta_{ij}(\cdot)$, $x_i(\cdot)$ are now defined by
\begin{align}
&\bar{\Delta}_{ij}(p,\xi_n)=\frac{p_j(p,\xi_n)\epsilon_i(\imath\xi_n)-p_i(p,\xi_n)\epsilon_j(\imath\xi_n)}{p_j(p,\xi_n)\epsilon_i(\imath\xi_n)+p_i(p,\xi_n)\epsilon_j(\imath\xi_n)}\,,\label{Hamakerdipole2}\\
&\Delta_{ij}(p,\xi_n)=\frac{p_j(p,\xi_n)-p_i(p,\xi_n)}{p_j(p,\xi_n)+p_i(p,\xi_n)}\,,\label{Hamakerdipole3}\\
&p_i(p,\xi_n)=\sqrt{p^2-1+\frac{\epsilon_i(\imath\xi_n)}{\epsilon_3(\imath\xi_n)}}\,.\label{Hamakerdipole4}
\end{align}

Within the \emph{low temperature dipole approximation}\,\cite{Lifshit6,Lifshit7}, it is implicitly assumed that instantaneously induced dipoles of quantum nature provide the dominant contribution to the van der Waals interaction. In other words, thermal fluctuations as well as high-order multipoles are ignored, which implies that both the low temperature limit and the dipole limit are simultaneously applied to the general retarded Hamaker form. The most convenient starting point is the low temperature Hamaker form, see Eq.(\ref{Hamakerlowtemp1}), on which the mathematical steps of the dipole approximation derivation can be sequentially applied. This leads to the Hamaker coefficient
\begin{align}
A_{132}^{\mathrm{dp,lt}}(l)&=+\frac{3\hbar}{4\pi}\int_{\frac{2\pi{k}_{\mathrm{b}}T}{\hbar}}^{\infty}\int_{1}^{\infty}r^2(l,\xi)p\left[\bar{\Delta}_{13}(p,\xi)\bar{\Delta}_{23}(p,\xi)\right.\nonumber\\&\,\,\,\,\,\,\left.+\Delta_{13}(p,\xi)\Delta_{23}(p,\xi)\right]e^{-r(l,\xi)p}dpd\xi+\frac{3}{4}k_{\mathrm{b}}T\zeta(3)\nonumber\\&\,\,\,\,\,\,\,+\frac{3}{4}k_{\mathrm{b}}T\mathrm{Li}_3\left(\Delta_{13}^{\mathrm{s}}\Delta_{23}^{\mathrm{s}}\right)\,,\label{Hamakerdipolelowtemp1}
\end{align}
where $\Delta_{ij}^{\mathrm{s}}$ and $r(\cdot)$ are still described by Eq.(\ref{Hamakergeneral2}) and Eq.(\ref{Hamakerlowtemp5}), respectively, while the auxiliary functions $\bar{\Delta}_{ij}(\cdot)$,\,$\Delta_{ij}(\cdot)$,\,$x_i(\cdot)$ are now defined by
\begin{align}
&\bar{\Delta}_{ij}(p,\xi)=\frac{p_j(p,\xi)\epsilon_i(\imath\xi)-p_i(p,\xi)\epsilon_j(\imath\xi)}{p_j(p,\xi)\epsilon_i(\imath\xi)+p_i(p,\xi)\epsilon_j(\imath\xi)}\,,\label{Hamakerdipolelowtemp2}\\
&\Delta_{ij}(p,\xi)=\frac{p_j(p,\xi)-p_i(p,\xi)}{p_j(p,\xi)+p_i(p,\xi)}\,,\label{Hamakerdipolelowtemp3}\\
&p_i(p,\xi)=\sqrt{p^2-1+\frac{\epsilon_i(\imath\xi)}{\epsilon_3(\imath\xi)}}\,.\label{Hamakerdipolelowtemp4}
\end{align}

Finally, within the \emph{modified low temperature dipole approximation}\,\cite{Lifshit6,Lifshit7}, the low temperature replacement of the discrete frequency spectrum with the continuous frequency spectrum is applied from the dc frequency rather than from the first bosonic Matsubara frequency, while the dipole limit is applied as per usual. Overall, these lead to the Hamaker coefficient
\begin{align}
A_{132}^{\mathrm{dp,lt,mod}}(l)&=+\frac{3\hbar}{4\pi}\int_{0}^{\infty}\int_{1}^{\infty}r^2(l,\xi)p\left[\bar{\Delta}_{13}(p,\xi)\bar{\Delta}_{23}(p,\xi)\right.\nonumber\\&\,\,\,\,\,\,\left.+\Delta_{13}(p,\xi)\Delta_{23}(p,\xi)\right]e^{-r(l,\xi)p}dpd\xi\nonumber\\&\,\,\,\,\,\,\,+\frac{3}{4}k_{\mathrm{b}}T\mathrm{Li}_3\left(\Delta_{13}^{\mathrm{s}}\Delta_{23}^{\mathrm{s}}\right)\,,\label{Hamakerdipolelowtempmod1}
\end{align}
where $\Delta_{ij}^{\mathrm{s}}$ and $r(\cdot)$ are still described by Eq.(\ref{Hamakergeneral2}) and Eq.(\ref{Hamakerlowtemp5}), respectively, while the $\bar{\Delta}_{ij}(\cdot)$,\,$\Delta_{ij}(\cdot)$,\,$x_i(\cdot)$ are still described by Eqs.(\ref{Hamakerdipolelowtemp2},\ref{Hamakerdipolelowtemp3},\ref{Hamakerdipolelowtemp4}), respectively.

\section{Computational aspects}\label{computational}

\subsection{Computation of the dielectric function at imaginary arguments}\label{ComputationalD}

\noindent In full Lifshitz theory as well as in its approximations, the computation of the retarded Hamaker coefficient requires the evaluation of the dielectric response at the imaginary bosonic Matsubara frequencies, \emph{i.e.}, $\epsilon(\imath\xi_n)$. Two alternative methods have been developed for the computation of $\epsilon(\imath\xi_n)$ from available experimental data of the complex dielectric function; the full spectral method and the simple spectral method. The first method is more accurate, but is only applicable when dense extended-in-frequency dielectric data are available; it will be utilized for metallic media. The second method is tedious and rather subjective, but is also applicable when sparse or limited-in-frequency dielectric data are available; it will be utilized for pure water.

For the $26$ elemental polycrystalline metals of interest, extended-in-frequency dielectric data up to $\hbar\omega=10000\,$eV are available\,\cite{opticalB}. Therefore, the \emph{full spectral method}, that is based on the Kramers-Kronig causality relations\,\cite{Lifshit1,Lifshit2}, will be followed for the determination of the dielectric function at the imaginary arguments\,\cite{dielect1,dielect2,dielect3,dielect4}. The basic expression reads as\,\cite{LandauB1,LandauB2}
\begin{equation}
\epsilon(\imath\xi_n)=1+\frac{2}{\pi}\int_0^{\infty}\frac{\omega\Im\{\epsilon(\omega)\}}{\omega^2+\xi_n^2}d\omega\,.\label{KramersKronig}
\end{equation}
The full spectral method only requires experimental data for the imaginary part of the dielectric function to compute $\epsilon(\imath\xi_n)$. It is important to note that $\epsilon(\imath\omega)$ is a monotonically decreasing function of the frequency that tends to infinity at zero frequencies for perfectly conducting media and that tends to unity asymptotically for any medium\,\cite{intromin}. It is a surprisingly structure-less function of the frequency, especially when compared to the notoriously non-monotonic $\Im\{\epsilon(\omega)\}$ and $\Re\{\epsilon(\omega)\}$ functions which contain multiple resonance signatures\,\cite{dielect3}.

On the other hand, for pure room temperature water, dielectric data up to $25\,$eV have been traditionally considered\,\cite{Lifshit5,waterdi1,waterdi2} and dielectric data up to $100\,$eV were only rather recently made available\,\cite{waterdi3,waterdi4,waterdi5}. In spite of arguments that the recent dielectric data do not require any artificial extrapolations for $\Im\{\epsilon(\omega)\}$ beyond $100\,$eV for the accurate computation of $\epsilon(\imath\xi_n)$\,\cite{waterdi6}, the data are judged to be insufficiently extended towards high frequencies for an unbiased application of the Kramers-Kronig relation, since all dielectric function related quantities are close but not at their asymptotic limits. The \emph{simple spectral method}, based on analytical parameterizations of the dielectric function, will be followed for the determination of the dielectric function at the imaginary arguments\,\cite{dielect3,waterdi7,waterdi8}. To be more specific, the model dielectric function universally assumed for water combines multiple Debye relaxation terms with a large number of Lorentz oscillators and reads as\,\cite{waterdi1,waterdi2,waterdi5}
\begin{align}
\epsilon(\omega)=1+\sum_j^{N_{\mathrm{D}}}\frac{c_j}{1-\imath\omega\tau_j}+\sum_j^{N_{\mathrm{L}}}\frac{d_j\omega_j^2}{\omega_j^2-\imath\omega\gamma_j-\omega^2}\,,\label{modeldielectric}
\end{align}
where $\tau_j$ is the Debye relaxation time, $c_j$ the polarization strength, $N_{\mathrm{D}}$ the Debye relaxation number, $\omega_j$ the resonant frequency, $\gamma_j$ the damping constant, $d_j$ the oscillator strength and $N_{\mathrm{L}}$ the Lorentz oscillator number. Once the necessary number of relaxation and oscillator terms is decided, the remaining unknown parameters are determined by simultaneous fits to experimental data for the real and imaginary parts of the dielectric function. Courtesy of analytic continuation, $\epsilon(\imath\xi_n)$ is ultimately obtained by a direct $\omega\to\imath\xi_n$ substitution in the model dielectric function.

\subsection{Computational input}\label{ComputationalI}

\noindent The necessary material input for the computation of the room temperature Hamaker coefficients concerns the long wavelength dielectric response and the relative static magnetic permeability of the interacting bodies (metals) and of the intervening medium (water).

\emph{Dielectric data of elemental metals.} Experimental room temperature dielectric data are adopted from Adachi's extended compilation\,\cite{opticalB} that contains systematic tabulations of the optical properties (complex dielectric function, complex refractive index, absorption coefficient and normal incidence reflectivity) of $63$ elemental metals as a function of the frequency. From these $63$ datasets, $11$ datasets concerned exclusively monocrystalline (anisotropic) metals and $26$ datasets were deemed to be inappropriate for accurate Lifshitz calculations owing to the presence of rather extended frequency gaps, the near absence of low frequency measurements at the infrared range and the sparsity of visible or ultraviolet measurements\,\cite{intromin}. Therefore, appropriate room temperature dielectric data were ultimately available for $26$ elemental polycrystalline metals spanning from the far infra-red to the soft X-ray region of the electromagnetic spectrum, roughly $50\,$meV-$10\,$keV. The $26$ selected elements include the most abundant transition metals (3d, 4d, 5d) as well as alkaline earth metals and even lanthanides. In these Adachi datasets, the highest photon frequency that is available corresponds to $\hbar\omega=10000\,$eV for all metals, whereas the lowest photon frequency that is available varies from $\hbar\omega=0.0062\,$eV up to $\hbar\omega=0.300\,$eV. In addition, the number of optical data points per element ranges from $254$ up to $495$.

\emph{Dielectric representation for pure water.} Fortunately, various parameterizations of the room temperature dielectric function of pure water are available in the literature, which spares us from the rather cumbersome task of curve fitting the respective data to the sum of Debye relaxation terms and of Lorentz oscillators, see Eq.(\ref{modeldielectric}). Owing to the current lack of consensus in the literature, three available parameterizations will be investigated herein. The classic Parsegian-Weiss representation that utilizes optical data within the $0-25\,$eV interval and employs one Debye term in the microwave range, five Lorenz terms in the infra-red range and six Lorenz terms in the ultraviolet range\,\cite{Lifshit5,waterdi1}. The standard Roth-Lenhoff representation that utilizes the same $0-25\,$eV optical measurements and employs the same Debye term in the microwave range, the same five Lorenz terms in the infra-red range and six updated Lorenz terms in the ultraviolet range\,\cite{waterdi2}. The contemporary representation of Fiedler \emph{et al.} that utilizes extended optical data within the $0-100\,$eV interval and employs two Debye terms in the microwave range, seven Lorenz terms in the infra-red range and twelve Lorenz terms in the ultraviolet range\,\cite{waterdi5}. For completeness, for all three parameterizations, the fitting parameters of the Debye terms are provided in Table \ref{dielectricDebye} and the fitting parameters of the Lorentz terms are provided in Table \ref{dielectricLorentz}.

\emph{Volume magnetic susceptibilities.} In case of ferromagnetic metals, the maximum relative magnetic permeability of high purity samples is employed ($\mu=100000$ for iron, $\mu=600$ for nickel, $\mu=250$ for cobalt\,\cite{intromin}) and converted to the volume magnetic susceptibility. In case of paramagnetic or diamagnetic metals; the molar magnetic susceptibility is employed, as tabulated in the CRC handbook\,\cite{magneti4}, and converted to the volume magnetic susceptibility.

\begin{table}[!b]
\centering
\caption{Fitting parameters of the room temperature dielectric function of pure water within the Parsegian-Weiss representation\,\cite{waterdi1}, Roth-Lenhoff representation\,\cite{waterdi2} \& Fiedler \emph{et al.} representation\,\cite{waterdi5}: the coefficients $1/\tau_j$, $c_j$, of the Debye relaxation terms, see Eq.(\ref{modeldielectric}). Notice that the Parsegian-Weiss fitting parameters are adopted from Parsegian's handbook\,\cite{Lifshit5}, where their values are updated compared to the original Parsegian and Weiss paper\,\cite{waterdi1}.}\label{dielectricDebye}
\begin{tabular}{c c}
 \hline\hline
$1/\tau_j$ (eV)      & $c_j$                                              \\ \hline
\multicolumn{2}{c}{Parsegian - Weiss representation\,\cite{waterdi1}}     \\
$6.55\times10^{-5}$  & $74.8$                                             \\ \hline
\multicolumn{2}{c}{Roth - Lenhoff representation\,\cite{waterdi2}}        \\
$6.50\times10^{-5}$  & $75.38$                                            \\ \hline
\multicolumn{2}{c}{Fiedler \emph{et al.} representation\,\cite{waterdi5}} \\
$6.84\times10^{-6}$	 & 0.47	                                              \\
$7.98\times10^{-5}$	 & 72.62                                              \\ \hline \hline
\end{tabular}
\end{table}

\begin{table}[!t]
\centering
\caption{Fitting parameters of the room temperature dielectric function of pure water within the Parsegian-Weiss representation\,\cite{waterdi1}, Roth-Lenhoff representation\,\cite{waterdi2} \& Fiedler \emph{et al.} representation\,\cite{waterdi5}: the coefficients $\omega_j$, $d_j$, $\gamma_j$ of the Lorentz oscillator terms, see Eq.(\ref{modeldielectric}). Note that the Parsegian-Weiss and the Roth-Lenhoff representations provide a dimensional oscillator strength $g_j$ that has been converted to the dimensionless oscillator strength $d_j=g_j/\omega_j^2$ of our notation. Notice also that the Parsegian-Weiss fitting parameters are adopted from Parsegian's handbook\,\cite{Lifshit5}, where their values are slightly updated compared to the original Parsegian and Weiss paper\,\cite{waterdi1}.}\label{dielectricLorentz}
\begin{tabular}{c c c}
 \hline\hline
$\omega_j$ (eV)     & $d_j$               & $\gamma_j$ (eV)                 \\ \hline
\multicolumn{3}{c}{Parsegian - Weiss representation\,\cite{waterdi1}}       \\
$2.07\times10^{-2}$ & $1.46$              & $1.5\times10^{-2}$              \\
$6.9\times10^{-2}$  & $7.35\times10^{-1}$ & $3.8\times10^{-2}$              \\	
$9.2\times10^{-2}$  & $1.51\times10^{-1}$ & $2.8\times10^{-2}$              \\	
$2.0\times10^{-1}$  & $1.36\times10^{-2}$ & $2.5\times10^{-2}$              \\	
$4.2\times10^{-1}$  & $7.65\times10^{-2}$ & $5.6\times10^{-2}$              \\
$8.25$              & $3.94\times10^{-2}$ & $0.51$                          \\		
$10.0$              & $5.67\times10^{-2}$ & $0.88$                          \\			
$11.4$              & $9.23\times10^{-2}$ & $1.54$                          \\
$13.0$              & $1.56\times10^{-1}$ & $2.05$                          \\			
$14.9$              & $1.52\times10^{-1}$ & $2.96$                          \\
$18.5$              & $2.71\times10^{-1}$ & $6.26$                          \\ \hline
\multicolumn{3}{c}{Roth - Lenhoff representation\,\cite{waterdi2}}          \\
$2.1\times10^{-2}$  & $1.43$              & $1.5\times10^{-2}$              \\
$6.9\times10^{-2}$  & $7.35\times10^{-1}$ & $3.8\times10^{-2}$              \\	
$9.2\times10^{-2}$  & $1.54\times10^{-1}$ & $2.8\times10^{-2}$              \\	
$2.0\times10^{-1}$  & $1.35\times10^{-2}$ & $2.5\times10^{-2}$              \\	
$4.2\times10^{-1}$  & $7.94\times10^{-2}$ & $5.6\times10^{-2}$              \\
$8.21$              & $4.84\times10^{-2}$ & $0.63$                          \\		
$10.0$              & $3.87\times10^{-2}$ & $0.84$                          \\			
$11.4$              & $9.23\times10^{-2}$ & $2.05$                          \\
$13.6$              & $3.44\times10^{-1}$ & $3.90$                          \\			
$17.8$              & $3.60\times10^{-1}$ & $7.33$                          \\
$25.2$              & $3.83\times10^{-2}$ & $5.34$                          \\ \hline
\multicolumn{3}{c}{Fiedler \emph{et al.} representation\,\cite{waterdi5}}   \\
$8.46\times10^{-4}$ & $2.59\times10^{-1}$ & $3.92\times10^{-4}$             \\
$4.19\times10^{-3}$ & $1.04$              & $7.43\times10^{-3}$             \\	
$2.12\times10^{-2}$ & $1.62$              & $2.60\times10^{-2}$             \\	
$6.25\times10^{-2}$ & $5.55\times10^{-1}$ & $3.98\times10^{-2}$             \\	
$8.49\times10^{-2}$ & $2.38\times10^{-1}$ & $2.99\times10^{-2}$             \\
$2.04\times10^{-1}$ & $1.34\times10^{-2}$ & $8.43\times10^{-3}$             \\		
$4.18\times10^{-1}$ & $7.17\times10^{-2}$ & $3.41\times10^{-2}$             \\			
$8.34$              & $4.47\times10^{-2}$ & $0.75$                          \\
$9.50$              & $3.27\times10^{-2}$ & $1.12$                          \\				
$10.41$             & $4.66\times10^{-2}$ & $1.26$                          \\
$11.67$             & $6.67\times10^{-2}$ & $1.58$                          \\	
$12.95$             & $7.42\times10^{-2}$ & $1.65$                          \\
$14.13$             & $9.30\times10^{-2}$ & $1.86$                          \\	
$15.50$             & $7.79\times10^{-2}$ & $2.22$                          \\	
$17.17$             & $7.90\times10^{-2}$ & $2.70$                          \\	
$18.89$             & $4.18\times10^{-2}$ & $2.82$                          \\	
$21.45$             & $1.07\times10^{-1}$ & $6.87$                          \\
$30.06$             & $1.33\times10^{-1}$ & $18.28$                         \\
$49.45$             & $5.66\times10^{-2}$ & $36.28$                         \\ \hline \hline
\end{tabular}
\end{table}

\subsection{Computational details}\label{ComputationalA}

\noindent Lifshitz theory calculations of the retarded Hamaker coefficient, either with the full theory or with approximations, require the numerical computation of multiple single or double improper integrals as well as of the Matsubara series. The infinite extent of these mathematical operations necessitates a number of truncations and extrapolations that should be made in a manner that neither blows up the computational cost nor compromises the accuracy.

\emph{Numerical quadrature}. Single and double integrations are carried out with the Gauss-Kronrod adaptive method. The $dx$ integrals in the interval $[r,\infty)$ and the $dp$ integrals in the interval $[1,\infty)$ do not require the introduction of extrapolations or of artificial cut-offs. On the other hand, the improper integrals over the frequency, see the $d\xi,d\omega$ integrals in Eqs.(\ref{Hamakerlowtemp1},\ref{Hamakerlowtempmod1},\ref{Hamakerdipolelowtemp1},\ref{Hamakerdipolelowtempmod1},\ref{KramersKronig}), require knowledge of the imaginary argument dielectric function or the imaginary part of the dielectric function at all frequencies. Since the available dielectric data for metals extend to high enough frequencies where $\epsilon(\imath\omega)$, $\Im\{\epsilon(\omega)\}$ have truly reached their asymptotic limits, upper frequency extrapolations are not necessary. Therefore, an upper integration limit of $\hbar\omega_{\mathrm{u}}=10000\,$eV can be directly imposed. Furthermore, the improper integrals that extend over the entire frequency domain, see the $d\xi,d\omega$ integrals in Eqs.(\ref{Hamakerlowtempmod1},\ref{Hamakerdipolelowtempmod1},\ref{KramersKronig}), require knowledge of the imaginary argument dielectric function or the imaginary part of the dielectric function at very low frequencies. In fact, given that $\epsilon(0)\to\infty$ for perfectly conducting media, the Hamaker coefficient can be rather sensitive to the lower integration limit and extrapolation strategies down to zero frequencies are necessary. In this work, a Drude extrapolation procedure will be followed, since the free electron intra-band effects can be expected to dominate the low frequency response at frequencies lower than the resonances associated with bound electron inter-band effects\,\cite{intromin}. In particular, a Drude model is assumed to be valid at the extrapolated range below the lowest frequency measurement $\hbar\omega_{\mathrm{l}}$, whose plasma frequency $\omega_{\mathrm{p}}$ and damping constant $\Gamma$ are determined by least-square fitting to the low frequency data $\hbar\omega\leq0.6$eV. Note that this fitting range contains $4$ to $69$ points in the Adachi datasets, depending on the element. The above lead to the equivalent Kramers-Kronig expression\,\cite{intromin}
\begin{equation}
\epsilon(\imath\xi_n)=\epsilon_{\mathrm{D}}(\imath\xi_n)+\frac{2}{\pi}\int_{\omega_{\mathrm{l}}}^{\omega_{\mathrm{u}}}\frac{\omega\left[\Im\{\epsilon(\omega)-\epsilon_{\mathrm{D}}(\omega)\}\right]}{\omega^2+\xi_n^2}d\omega\,,\label{KramersKronigNew}
\end{equation}
where $\epsilon_{\mathrm{D}}(\omega)=1-\omega_{\mathrm{p}}^2/(\omega^2+\imath\omega\Gamma)$ for the Drude model dielectric function. Finally, concerning the application of the Kramers-Kronig expression, it is also worth noting that Hermite polynomial interpolation schemes are utilized under the positivity constraint in order to construct an analytic $\Im\{\epsilon(\omega)\}$ representation, so that $\Im\{\epsilon(\omega)\}$ can be readily computed at all evaluation points of the Gauss-Kronrod algorithm.

\emph{Numerical summation}. The bosonic Matsubara series allows the computation of the additive contributions to the retarded Hamaker coefficient that originate from different electromagnetic frequency ($\omega$) ranges or, equivalently, different virtual photon energy ($\hbar\omega$) ranges. For a constant separation $l$, the $\xi_n-$dependence of the auxiliary functions $\bar{\Delta}_{ij}(l,x,\xi_n)$, $\Delta_{ij}(l,x,\xi_n)$ in connection with the monotonic $\epsilon(\imath\omega)$ decrease with increasing frequency imply that contributions from increasing Matsubara frequencies gradually decrease. In the present calculations, at short separations $l\lesssim5\,$nm, the Matsubara series is truncated at $n=61564$, which corresponds to the last Matsubara frequency prior to $10000\,$eV. The residual contributions from all neglected Matsubara orders are expected to be at least six orders of magnitude lower, as deduced from our earlier investigation of the non-retarded limit\,\cite{intromin}. As a consequence of the well-documented relativistic suppression of high frequency interactions\,\cite{Lifshit5}, the Matsubara series is truncated at progressively lower $n$ as the separation increases. For instance, for separations $l\sim30\,$nm the Matsubara series can be truncated with negligible errors at $n=1846$, which corresponds to the last Matsubara frequency prior to $300\,$eV, while for separations $l\sim200\,$nm the Matsubara series can be truncated with negligible errors at $n=153$, which corresponds to the last Matsubara frequency prior to $25\,$eV. It is evident that our adaptive control of the Matsubara summation cut-off leads to a drastic reduction in the computational cost.

\begin{table}[!t]
\centering
\caption{A detailed comparison of the room temperature Hamaker coefficients between $26$ identical elemental polycrystalline metals that are embedded in vacuum; as computed from the full Lifshitz theory [see Eqs.(\ref{Hamakergeneral1},\ref{Hamakergeneral2},\ref{Hamakergeneral3},\ref{Hamakergeneral4},\ref{Hamakergeneral5},\ref{Hamakergeneral6})], low temperature approximation [see Eqs.(\ref{Hamakergeneral2},\ref{Hamakerlowtemp1},\ref{Hamakerlowtemp2},\ref{Hamakerlowtemp3},\ref{Hamakerlowtemp4},\ref{Hamakerlowtemp5}), superscript \enquote{lt}], modified low temperature approximation [see Eqs.(\ref{Hamakergeneral2},\ref{Hamakerlowtemp2},\ref{Hamakerlowtemp3},\ref{Hamakerlowtemp4},\ref{Hamakerlowtemp5},\ref{Hamakerlowtempmod1}), superscript \enquote{lt,mod}], dipole approximation [see Eqs.(\ref{Hamakergeneral2},\ref{Hamakergeneral6},\ref{Hamakerdipole1},\ref{Hamakerdipole2},\ref{Hamakerdipole3},\ref{Hamakerdipole4}), superscript \enquote{dp}], low temperature dipole approximation [see Eqs.(\ref{Hamakergeneral2},\ref{Hamakerlowtemp5},\ref{Hamakerdipolelowtemp1},\ref{Hamakerdipolelowtemp2},\ref{Hamakerdipolelowtemp3},\ref{Hamakerdipolelowtemp4}), superscript \enquote{dp,lt}] and modified low temperature dipole approximation [see Eqs.(\ref{Hamakergeneral2},\ref{Hamakerlowtemp5},\ref{Hamakerdipolelowtemp2},\ref{Hamakerdipolelowtemp3},\ref{Hamakerdipolelowtemp4},\ref{Hamakerdipolelowtempmod1}), superscript \enquote{dp,lt,mod}]. The absolute relative errors $e_{\mathrm{r}}$ between each approximation and the full theory are reported for each element, after averaging within the separation interval $l=0-100\,$nm.}\label{hamakernanocomparison}
\begin{tabular}{c c c c c c c c}
\\ \hline\hline
      &	$e_{\mathrm{r}}^{\mathrm{lt}}$ & $e_{\mathrm{r}}^{\mathrm{lt,mod}}$ & $e_{\mathrm{r}}^{\mathrm{dp}}$ & $e_{\mathrm{r}}^{\mathrm{dp,lt}}$ & $e_{\mathrm{r}}^{\mathrm{dp,lt,mod}}$   \\
	  &  $\%$                          & $\%$                               & $\%$                           & $\%$                              & $\%$                                    \\ \hline\hline
Ag    &  $3.79$                        & $0.42$                             & $6.81$                         &  $10.44$                          & $6.99$                                  \\	
Al    &  $3.23$                        & $0.53$                             & $7.77$                         &  $10.89$                          & $7.72$                                  \\	
Au    &  $3.64$                        & $0.34$                             & $6.57$                         &  $10.04$                          & $6.82$                                  \\	
Ba    &  $4.41$                        & $0.05$                             & $6.41$                         &  $10.62$                          & $7.15$                                  \\	
Be    &  $3.32$                        & $0.32$                             & $6.92$                         &  $10.10$                          & $7.14$                                  \\	
Co    &  $3.21$                        & $0.13$                             & $6.00$                         &  $9.03\,\,$                       & $6.41$                                  \\	
Cr    &  $3.39$                        & $0.30$                             & $6.64$                         &  $9.88\,\,$                       & $6.90$                                  \\	
Cu    &  $3.79$                        & $0.32$                             & $6.61$                         &  $10.24$                          & $6.91$                                  \\	
Fe    &  $3.31$                        & $0.32$                             & $6.25$                         &  $9.39\,\,$                       & $6.49$                                  \\	
Hf    &  $3.94$                        & $0.07$                             & $5.21$                         &  $8.96\,\,$                       & $6.02$                                  \\	
Ir    &  $2.96$                        & $0.24$                             & $6.82$                         &  $9.66\,\,$                       & $7.05$                                  \\	
Mo    &  $3.19$                        & $0.45$                             & $6.90$                         &  $9.95\,\,$                       & $6.95$                                  \\	
Nb    &  $3.26$                        & $0.31$                             & $6.63$                         &  $9.75\,\,$                       & $6.85$                                  \\	
Ni    &  $3.35$                        & $0.14$                             & $6.06$                         &  $9.24\,\,$                       & $6.50$                                  \\	
Os    &  $3.15$                        & $0.05$                             & $5.73$                         &  $8.72\,\,$                       & $6.24$                                  \\	
Pd    &  $3.40$                        & $0.05$                             & $6.00$                         &  $9.24\,\,$                       & $6.55$                                  \\	
Pt    &  $3.25$                        & $0.18$                             & $6.33$                         &  $9.43\,\,$                       & $6.70$                                  \\	
Rh    &  $3.15$                        & $0.26$                             & $6.80$                         &  $9.81\,\,$                       & $7.05$                                  \\	
Sc    &  $3.96$                        & $0.02$                             & $5.87$                         &  $9.64\,\,$                       & $6.59$                                  \\	
Sr    &  $4.19$                        & $0.09$                             & $6.32$                         &  $10.32$                          & $6.97$                                  \\	
Ta    &  $3.37$                        & $0.38$                             & $6.71$                         &  $9.93\,\,$                       & $6.86$                                  \\	
Ti    &  $4.00$                        & $0.18$                             & $6.09$                         &  $9.89\,\,$                       & $6.61$                                  \\	
Tm    &  $3.45$                        & $0.11$                             & $6.22$                         &  $9.50\,\,$                       & $6.71$                                  \\	
V     &  $3.41$                        & $0.09$                             & $6.10$                         &  $9.34\,\,$                       & $6.61$                                  \\	
W     &  $3.19$                        & $0.27$                             & $6.32$                         &  $9.35\,\,$                       & $6.58$                                  \\	
Zr    &  $3.76$                        & $0.12$                             & $6.10$                         &  $9.67\,\,$                       & $6.64$                                  \\	\hline\hline
\end{tabular}
\end{table}

\section{Numerical results}\label{numerical}

\noindent As aforementioned, the Lifshitz computation of Hamaker coefficients as a function of the separation is sparse in the literature, regardless of material-medium-material combination\,\cite{Lifshit5,LifretP1,LifretP2,LifretP3,LifretP4}. Apart from the inherent numerical complexity of the full Lifshitz theory, another obstacle is the absence of a simple accurate expression for the dependence of the Hamaker coefficients on the separation $l$ that would substitute lengthy tabulations. Below, we shall devise such a convenient semi-empirical expression.

In general, the introduction of a characteristic absorption frequency $\omega_0$ facilitates the distinction of three asymptotic regimes, where the dispersion force reduces to simple power laws\,\cite{asympto1,asympto2,asympto3}. Within the small separation (non-retarded) limit, $l\ll{c}/\omega_0$, one retrieves the non-relativistic van der Waals force that reads as $F_{\mathrm{vdW}}=-A_0/(6\pi{l}^3)$ with $A_0$ the Hamaker constant\,\cite{Lifshit6,Lifshit2,Lifshit5}. Within the large separation (fully retarded) limit, ${c}/\omega_0\ll{l}\ll\hbar{c}/(k_{\mathrm{b}}T)$, one retrieves the Casimir force that has the well-known form $F_{\mathrm{Cas}}=-H/(6\pi{l}^4)$ with $H=(\pi^3/40)\hbar{c}\simeq24506.8\,$zJ$\cdot$nm a material independent constant\,\cite{Lifshit2,Lifshit5}. Within the very large separation (thermal) limit, ${l}\gg\hbar{c}/(k_{\mathrm{b}}T)$, one obtains the dispersion force $F_{\mathrm{th}}=-A_{\mathrm{th}}/(6\pi{l}^3)$ where $A_{\mathrm{th}}=(3/4)k_{\mathrm{b}}T[\zeta(3)+\mathrm{Li}_3(\Delta_{13}^{\mathrm{s}}\Delta_{23}^{\mathrm{s}})]$ denotes the static magneto-dielectric contribution. For metals, the latter limit is realized in the micrometric range, \emph{i.e.}, at separations far larger than those of interest herein.

After translating to the Hamaker coefficient\,\cite{HamakerP}, the former two asymptotic limits imply a Hamaker constant $\propto{l}^{0}$ as $l\to0$ and also a Hamaker function $\propto{l}^{-1}$ as $l\to\infty$. The most straightforward way to capture such a functional behavior is by considering a simple fraction $1/(a+bl)$. Given the complicated extended-in-frequency optical response of metals, we shall adopt a sum of two simple fractions, \emph{i.e.}
\begin{align}
A(l)=\frac{1}{a+bl}+\frac{1}{c+dl}\,,\label{hamakerfit}
\end{align}
with the unknown material dependent parameters $a,b,c,d$ determined by least square fits to the computed Hamaker coefficients. Since it is customary to express the Hamaker coefficients $A$ in zJ and the semi-space separations $l$ in nm, then $a,c$ will be typically expressed in (zJ)$^{-1}$ units and $b,d$ typically expressed in (zJ$\cdot$nm)$^{-1}$ units. It should be pointed out that the exact asymptotic limits generate two rigorous constraints for the $a,b,c,d$ parameters:
\begin{align}
&A_0=\frac{1}{a}+\frac{1}{c}\,,\label{hamakerconstrained1}\\
&H=\frac{1}{b}+\frac{1}{d}\,.\label{hamakerconstrained2}
\end{align}
The unconstrained and the doubly-constrained versions of the semi-empirical expression will both be considered.

In this section, we shall exclusively focus on submicron separations $l\leq1\,\mu$m and mainly on separations $l\leq200\,$nm. At such separations, retarded van der Waals forces between metals are more likely to be important in practical applications. The employed discretization scheme with respect to the separation of the metallic semi-spaces is the following: $l=0-1\,$nm \& $\Delta{l}=0.1\,$nm which corresponds to $11$ data points, $l=1-200\,$nm \& $\Delta{l}=1\,$nm which corresponds to $200$ data points and $l=0.2-1\,\mu$m \& $\Delta{l}=5\,$nm which corresponds to $160$ data points.

\begin{table}[!t]
\centering
\caption{Parameterizations of the room temperature Hamaker coefficients between $26$ identical elemental polycrystalline metals that are embedded in vacuum, valid for separations within $0-200\,$nm, as computed from the full Lifshitz theory [see Eqs.(\ref{Hamakergeneral1},\ref{Hamakergeneral2},\ref{Hamakergeneral3},\ref{Hamakergeneral4},\ref{Hamakergeneral5},\ref{Hamakergeneral6})]. The $4$ coefficients $a,\,b,\,c,\,d$ of Eq.(\ref{hamakerfit}) are determined by least square fitting for separations within $0-200\,$nm \textbf{without constraints}. The mean absolute relative errors of the fits are also reported for each element.}\label{hamakernanovacuumNO}
\begin{tabular}{c c c c c c c c}
\\ \hline\hline
      &	       a           &     b	              &    c        &     d		           &  $e_{\mathrm{r}}$   \\
	  &  (zJ)$^{-1}$       & (zJ$\cdot$nm)$^{-1}$ & (zJ)$^{-1}$ & (zJ$\cdot$nm)$^{-1}$ &  $\%$               \\ \hline\hline
Ag    &  $0.0117\,\,\,$    &  $0.0000782$  		  &  $0.00348$  &  $0.000354$  		   &  $0.27$             \\
Al	  &  $0.00530$         &  $0.0000781$  		  &  $0.00586$  &  $0.000404$ 		   &  $0.35$             \\
Au	  &  $0.0142\,\,\,$    &  $0.0000864$  		  &  $0.00297$  &  $0.000280$ 		   &  $0.20$             \\
Ba	  &  $0.0254\,\,\,$    &  $0.000120\,\,\,$    &  $0.00703$  &  $0.000247$ 		   &  $0.16$             \\
Be	  &  $0.0178\,\,\,$    &  $0.000114\,\,\,$    &  $0.00329$  &  $0.000173$ 		   &  $0.22$             \\
Co	  &  $0.0167\,\,\,$    &  $0.0000995$  		  &  $0.00278$  &  $0.000226$ 		   &  $0.13$             \\
Cr 	  &  $0.0134\,\,\,$    &  $0.000108\,\,\,$    &  $0.00332$  &  $0.000235$ 		   &  $0.12$             \\
Cu	  &  $0.0146\,\,\,$    &  $0.0000880$  		  &  $0.00365$  &  $0.000292$ 		   &  $0.15$             \\
Fe	  &  $0.0153\,\,\,$    &  $0.0000897$  		  &  $0.00302$  &  $0.000229$ 		   &  $0.10$             \\
Hf	  &  $0.0179\,\,\,$    &  $0.000169\,\,\,$    &  $0.00450$  &  $0.000349$ 		   &  $0.13$             \\
Ir	  &  $0.00887$         &  $0.000100\,\,\,$    &  $0.00233$  &  $0.000245$ 		   &  $0.21$             \\
Mo	  &  $0.0114\,\,\,$    &  $0.000100\,\,\,$    &  $0.00250$  &  $0.000212$ 		   &  $0.23$             \\
Nb	  &  $0.0143\,\,\,$    &  $0.000102\,\,\,$    &  $0.00250$  &  $0.000214$ 		   &  $0.19$             \\
Ni	  &  $0.0144\,\,\,$    &  $0.0000939$  		  &  $0.00326$  &  $0.000266$ 		   &  $0.16$             \\
Os	  &  $0.0184\,\,\,$    &  $0.000172\,\,\,$    &  $0.00232$  &  $0.000213$ 		   &  $0.21$             \\
Pd	  &  $0.0148\,\,\,$    &  $0.000124\,\,\,$    &  $0.00307$  &  $0.000279$ 		   &  $0.08$             \\
Pt	  &  $0.0132\,\,\,$    &  $0.000112\,\,\,$    &  $0.00264$  &  $0.000250$ 		   &  $0.10$             \\
Rh	  &  $0.00964$         &  $0.0000965$  		  &  $0.00279$  &  $0.000267$ 		   &  $0.23$             \\
Sc	  &  $0.0283\,\,\,$    &  $0.000158\,\,\,$    &  $0.00496$  &  $0.000226$ 		   &  $0.11$             \\
Sr	  &  $0.0255\,\,\,$    &  $0.000108\,\,\,$    &  $0.00548$  &  $0.000243$ 		   &  $0.17$             \\
Ta	  &  $0.0128\,\,\,$    &  $0.0000950$  		  &  $0.00265$  &  $0.000238$ 		   &  $0.15$             \\
Ti	  &  $0.0208\,\,\,$    &  $0.000120\,\,\,$    &  $0.00451$  &  $0.000266$ 		   &  $0.08$             \\
Tm	  &  $0.0203\,\,\,$    &  $0.000139\,\,\,$    &  $0.00351$  &  $0.000208$ 		   &  $0.10$             \\
V	  &  $0.0186\,\,\,$    &  $0.000140\,\,\,$    &  $0.00329$  &  $0.000222$ 		   &  $0.11$             \\
W	  &  $0.0164\,\,\,$    &  $0.000121\,\,\,$    &  $0.00220$  &  $0.000201$ 		   &  $0.11$             \\
Zr	  &  $0.0192\,\,\,$    &  $0.000116\,\,\,$    &  $0.00383$  &  $0.000262$ 		   &  $0.15$             \\ \hline\hline
\end{tabular}
\end{table}

\begin{table}[!t]
\centering
\caption{Parameterizations of the room temperature Hamaker coefficients between $26$ identical elemental polycrystalline metals that are embedded in vacuum, valid for separations within $0-200\,$nm, as computed from the full Lifshitz theory [see Eqs.(\ref{Hamakergeneral1},\ref{Hamakergeneral2},\ref{Hamakergeneral3},\ref{Hamakergeneral4},\ref{Hamakergeneral5},\ref{Hamakergeneral6})]. The $4$ coefficients $a,\,b,\,c,\,d$ of Eq.(\ref{hamakerfit}) are determined by least square fitting for separations within $0-200\,$nm \textbf{with two physical constraints} [see Eqs.(\ref{hamakerconstrained1},\ref{hamakerconstrained2})]. The mean absolute relative errors of the fits are also reported for each element.}\label{hamakernanovacuumWITH}
\begin{tabular}{c c c c c c c c}
\\ \hline\hline
      &	       a           &     b	              &    c        &     d		           &  $e_{\mathrm{r}}$   \\
	  &  (zJ)$^{-1}$       & (zJ$\cdot$nm)$^{-1}$ & (zJ)$^{-1}$ & (zJ$\cdot$nm)$^{-1}$ &  $\%$               \\ \hline\hline
Ag    &  $0.0195\,\,\,$    &  $0.0000487$  		  &  $0.00316$  &  $0.000253$  		   &  $2.09$             \\
Al	  &  $0.0314\,\,\,$    &  $0.0000644$  		  &  $0.00306$  &  $0.000111$ 		   &  $1.84$             \\
Au	  &  $0.0237\,\,\,$    &  $0.0000502$  		  &  $0.00278$  &  $0.000219$ 		   &  $1.75$             \\
Ba	  &  $0.0575\,\,\,$    &  $0.0000530$         &  $0.00615$  &  $0.000177$ 		   &  $0.74$             \\
Be	  &  $0.0412\,\,\,$    &  $0.0000586$         &  $0.00302$  &  $0.000134$ 		   &  $1.02$             \\
Co	  &  $0.0301\,\,\,$    &  $0.0000526$  		  &  $0.00262$  &  $0.000182$ 		   &  $1.33$             \\
Cr 	  &  $0.0330\,\,\,$    &  $0.0000543$         &  $0.00294$  &  $0.000164$ 		   &  $1.61$             \\
Cu	  &  $0.0256\,\,\,$    &  $0.0000502$  		  &  $0.00334$  &  $0.000218$ 		   &  $1.58$             \\
Fe	  &  $0.0266\,\,\,$    &  $0.0000526$  		  &  $0.00282$  &  $0.000181$ 		   &  $1.26$             \\
Hf	  &  $0.0580\,\,\,$    &  $0.0000497$         &  $0.00387$  &  $0.000229$ 		   &  $1.96$             \\
Ir	  &  $0.0239\,\,\,$    &  $0.0000548$         &  $0.00203$  &  $0.000160$ 		   &  $2.56$             \\
Mo	  &  $0.0268\,\,\,$    &  $0.0000557$         &  $0.00226$  &  $0.000153$ 		   &  $1.87$             \\
Nb	  &  $0.0286\,\,\,$    &  $0.0000540$         &  $0.00233$  &  $0.000167$ 		   &  $1.64$             \\
Ni	  &  $0.0268\,\,\,$    &  $0.0000513$  		  &  $0.00298$  &  $0.000200$ 		   &  $1.53$             \\
Os	  &  $0.0531\,\,\,$    &  $0.0000538$         &  $0.00217$  &  $0.000169$ 		   &  $1.66$             \\
Pd	  &  $0.0360\,\,\,$    &  $0.0000512$         &  $0.00277$  &  $0.000200$ 		   &  $2.09$             \\
Pt	  &  $0.0303\,\,\,$    &  $0.0000525$         &  $0.00240$  &  $0.000184$ 		   &  $2.09$             \\
Rh	  &  $0.0242\,\,\,$    &  $0.0000535$  		  &  $0.00241$  &  $0.000172$ 		   &  $2.38$             \\
Sc	  &  $0.0699\,\,\,$    &  $0.0000531$         &  $0.00455$  &  $0.000176$ 		   &  $0.86$             \\
Sr	  &  $0.0467\,\,\,$    &  $0.0000519$         &  $0.00505$  &  $0.000191$ 		   &  $0.83$             \\
Ta	  &  $0.0253\,\,\,$    &  $0.0000528$  		  &  $0.00244$  &  $0.000179$ 		   &  $1.79$             \\
Ti	  &  $0.0441\,\,\,$    &  $0.0000513$         &  $0.00410$  &  $0.000199$ 		   &  $1.21$             \\
Tm	  &  $0.0512\,\,\,$    &  $0.0000548$         &  $0.00321$  &  $0.000160$ 		   &  $1.14$             \\
V	  &  $0.0469\,\,\,$    &  $0.0000537$         &  $0.00299$  &  $0.000170$ 		   &  $1.43$             \\
W	  &  $0.0348\,\,\,$    &  $0.0000544$         &  $0.00208$  &  $0.000163$ 		   &  $1.58$             \\
Zr	  &  $0.0397\,\,\,$    &  $0.0000513$         &  $0.00352$  &  $0.000199$ 		   &  $1.43$             \\ \hline\hline
\end{tabular}
\end{table}

\subsection{Identical metals in vacuum for $l\leq200\,$nm}\label{numericalvacuumshort}

\begin{figure}[!t]
	\centering
	\includegraphics[width=3.45in]{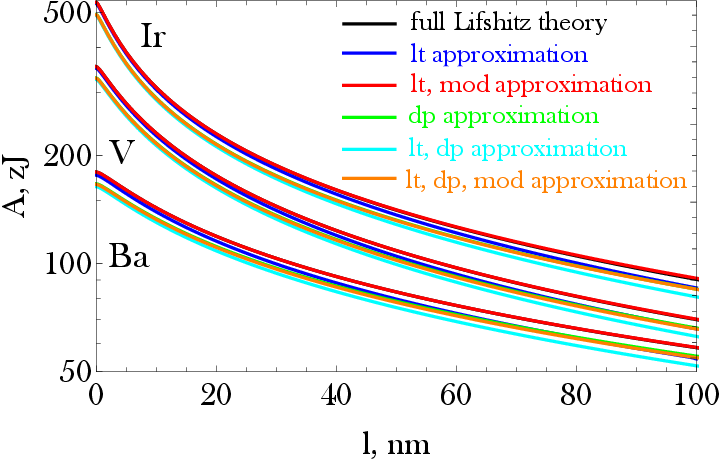}
	\caption{The room temperature Hamaker coefficients for Ir-Ir, V-V and Ba-Ba semi-spaces embedded in vacuum, computed from the full Lifshitz theory and its approximations, as functions of the separation.}\label{fig:Lifshitz_approximations}
\end{figure}

\noindent The retarded room temperature Hamaker coefficients of the van der Waals interactions between identical isotropic elemental metals that are embedded in vacuum have been computed from the full Lifshitz theory, low temperature approximation, modified low temperature approximation, dipole approximation, low temperature dipole approximation and modified low temperature dipole approximation. The level of accuracy of these five approximations has been quantified in Table \ref{hamakernanocomparison} at separations within $0-100\,$nm, for all the $26$ metals of interest. The following conclusions are due: \textbf{(i)} The modified version of the low temperature approximation is superior to the standard version of the low temperature approximation, \emph{i.e.} it is more accurate to consider the continuous spectrum from the zero frequency rather than from the first bosonic Matsubara frequency. \textbf{(ii)} Both versions of the low temperature approximation are superior to the dipole approximation, while the two combinations of both approximations are naturally characterized by the largest absolute errors. \textbf{(iii)} The accuracy level of any approximation only slightly depends on the metal composition. The rough element-averaged and separation-averaged errors are $3.50\%$ for the low temperature approximation, $0.22\%$ for the modified low temperature approximation, $6.39\%$ for the dipole approximation, $9.73\%$  for the low temperature dipole approximation and $6.77\%$ for the modified low temperature dipole approximation. \textbf{(iv)} The modified low temperature approximation is extremely accurate for all metals of interest. Its minimum, maximum and mean absolute relative deviations from the full Lifshitz theory results are $0.02\%$, $0.53\%$ and $0.22\%$, respectively. In Fig.\ref{fig:Lifshitz_approximations}, the room temperature Hamaker coefficients, as computed from the full Lifshitz theory and its approximations, are plotted as functions of the separation for three identical metal combinations that belong to different columns of the periodic table.

\begin{figure}[!t]
	\centering
	\includegraphics[width=3.45in]{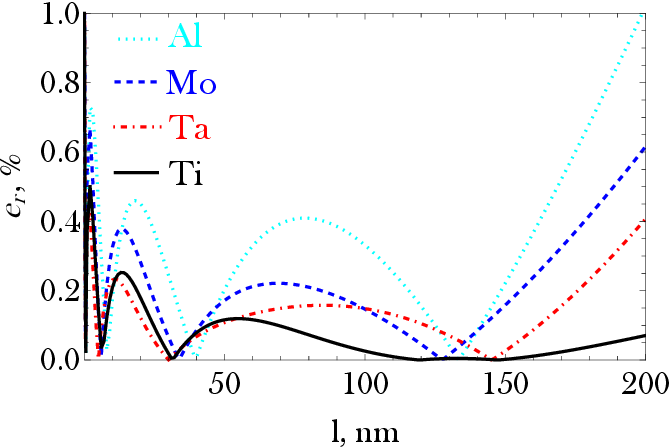}
	\caption{The absolute relative deviations between the Hamaker coefficients computed from full Lifshitz theory and the semi-empirical unconstrained Eq.(\ref{hamakerfit}) with least square parameters for Al-Al, Mo-Mo, Ta-Ta \& Ti-Ti semi-spaces embedded in vacuum, as functions of the separation.}\label{fig:fitting_errors_vacuum}
\end{figure}

Restricting the discussion to the exact Lifshitz results for the retarded room temperature Hamaker coefficients of identical polycrystalline elemental metals embedded in vacuum, the least-square fit parameters $a,b,c,d$ and the accuracy of the simple semi-empirical Eq.(\ref{hamakerfit}) are listed in Table \ref{hamakernanovacuumNO} and Table \ref{hamakernanovacuumWITH}. In particular, Table \ref{hamakernanovacuumNO} focuses on the unconstrained semi-empirical expression at separations $l\in[0,200]\,$nm, while Table \ref{hamakernanovacuumWITH} focuses on the doubly constrained semi-empirical expression, see Eqs.(\ref{hamakerconstrained1},\ref{hamakerconstrained2}), at separations $l\in[0,200]\,$nm. It is apparent that the unconstrained semi-empirical expression is extremely accurate within $[0,200]\,$nm regardless of the metal, with mean absolute relative errors ranging from $0.08\%$ (Pd, Ti) up to $0.35\%$ (Al) and an element-averaged accuracy of $0.16\%$. It is also evident that the doubly constrained semi-empirical expression is very accurate within $[0,200]\,$nm regardless of the metal, with mean absolute relative errors ranging from $0.74\%$ (Ba) up to $2.56\%$ (Ir) and an element-averaged accuracy of $1.59\%$. The incorporation of physical restrictions worsens the accuracy level, since the enforcement of exact constraints unavoidably leads to augmented relative errors at intermediate separations. It is also worth pointing out that the unconstrained least-square fitted parameters automatically satisfy the two physical restrictions to a very large degree. Moreover, in Fig.\ref{fig:fitting_errors_vacuum}, the absolute relative deviation between the unconstrained semi-empirical expression and the exact Lifshitz theory Hamaker coefficients is plotted as a function of the separation for identical metal combinations that belong to different columns of the periodic table. In spite of the non-monotonicity of the relative error, it is evident that there are no short separations $l\in[0,200]\,$nm that are accompanied by unacceptably large localized errors. This further demonstrates the very high quality of the proposed semi-empirical expression at short distances $l\in[0,200]\,$nm, regardless of the metal.

Finally, let us discuss the performance of the simplest single fraction form $A(l)=1/(a+bl)$. Its unconstrained version with least square fitted parameters $a,b$ is rather inaccurate with mean absolute relative errors ranging from $2.12\%$ (Al) up to $8.91\%$ (Au) and an element-averaged accuracy of $5.94\%$. Its doubly constrained version naturally leads to $a=A^{-1}$, $b=H^{-1}$ and does not require any least square fitting, but it is grossly inaccurate with mean absolute relative deviations ranging from $42.3\%$ (Al) up to $119.5\%$ (Hf) and an element-averaged accuracy of $82.8\%$. To be more specific, it leads to a drastic overestimation in the entire range of separations. In the literature, the unconstrained single fraction form $A(l)=H/[(H/A)+l]$ has been employed as a phenomenological model of the atom - wall dispersion interaction in the context of the quantum reflection of atoms from solid surfaces\,\cite{reflect1,reflect2,reflect3,reflect4}. Our results for the wall - wall dispersion interactions suggest that this is a very crude model even at a qualitative level, although it should be expected to be quite more accurate for atom - wall dispersion interactions given the much simpler absorption spectrum of atoms. This is supported by a dedicated study that revealed that the unconstrained single fraction form has an accuracy level of the order of $10\%$ for metastable helium atoms in the vicinity of gold and silicon walls\,\cite{reflect5}.

\subsection{Identical metals in vacuum for $l\leq1\,\mu$m}\label{numericalvacuumlong}

\noindent In what follows, we extend the discussion to the retarded room temperature Hamaker coefficients of identical polycrystalline elemental metals embedded in vacuum at separations $l\in[0,1]\,\mu$m. Table \ref{hamakermicronvacuum} contains the least-square fit parameters $a,b,c,d$ and the accuracy of the unconstrained semi-empirical Eq.(\ref{hamakerfit}).

\begin{table}[!t]
\centering
\caption{Parameterizations of the room temperature Hamaker coefficients between $26$ identical elemental polycrystalline metals that are embedded in vacuum, valid for separations within $0-1\,\mu$m, as computed from the full Lifshitz theory [see Eqs.(\ref{Hamakergeneral1},\ref{Hamakergeneral2},\ref{Hamakergeneral3},\ref{Hamakergeneral4},\ref{Hamakergeneral5},\ref{Hamakergeneral6})]. The $4$ coefficients $a,\,b,\,c,\,d$ of Eq.(\ref{hamakerfit}) are determined by least square fitting for separations within $0-1\,\mu$m \textbf{without constraints}. The mean absolute relative errors of the fits are also reported for each element.}\label{hamakermicronvacuum}
\begin{tabular}{c c c c c c c c}
\\ \hline\hline
      &	       a           &     b	              &    c        &     d		           &  $e_{\mathrm{r}}$   \\
	  &  (zJ)$^{-1}$       & (zJ$\cdot$nm)$^{-1}$ & (zJ)$^{-1}$ & (zJ$\cdot$nm)$^{-1}$ &  $\%$               \\ \hline\hline
Ag    &  $0.0101\,\,\,$    &  $0.0000842$  		  &  $0.00363$  &  $0.000403$  		   &  $2.55$             \\
Al	  &  $0.00421$         &  $0.0000748$  		  &  $0.00810$  &  $0.000748$ 		   &  $3.24$             \\
Au	  &  $0.0122\,\,\,$    &  $0.0000937$  		  &  $0.00307$  &  $0.000307$ 		   &  $2.04$             \\
Ba	  &  $0.0201\,\,\,$    &  $0.000124\,\,\,$	  &  $0.00757$  &  $0.000288$ 		   &  $0.62$             \\
Be	  &  $0.0130\,\,\,$    &  $0.000117\,\,\,$	  &  $0.00353$  &  $0.000201$ 		   &  $1.67$             \\
Co	  &  $0.0193\,\,\,$    &  $0.0000909$  		  &  $0.00272$  &  $0.000213$ 		   &  $1.01$             \\
Cr 	  &  $0.0118\,\,\,$    &  $0.000110\,\,\,$	  &  $0.00344$  &  $0.000255$ 		   &  $1.25$             \\
Cu	  &  $0.0125\,\,\,$    &  $0.0000941$  		  &  $0.00381$  &  $0.000324$ 		   &  $1.77$             \\
Fe	  &  $0.0169\,\,\,$    &  $0.0000849$  		  &  $0.00297$  &  $0.000218$ 		   &  $0.76$             \\
Hf	  &  $0.0208\,\,\,$    &  $0.000165\,\,\,$	  &  $0.00435$  &  $0.000322$ 		   &  $0.91$             \\
Ir	  &  $0.00787$         &  $0.000102\,\,\,$	  &  $0.00241$  &  $0.000265$ 		   &  $1.99$             \\
Mo	  &  $0.00957$         &  $0.000103\,\,\,$	  &  $0.00261$  &  $0.000234$ 		   &  $2.07$             \\
Nb	  &  $0.0122\,\,\,$    &  $0.000107\,\,\,$	  &  $0.00257$  &  $0.000231$ 		   &  $1.81$             \\
Ni	  &  $0.0168\,\,\,$    &  $0.0000862$  		  &  $0.00316$  &  $0.000245$ 		   &  $1.12$             \\
Os	  &  $0.0207\,\,\,$    &  $0.000166\,\,\,$	  &  $0.00229$  &  $0.000206$ 		   &  $0.56$             \\
Pd	  &  $0.0144\,\,\,$    &  $0.000125\,\,\,$	  &  $0.00309$  &  $0.000284$ 		   &  $0.59$             \\
Pt	  &  $0.0123\,\,\,$    &  $0.000114\,\,\,$	  &  $0.00267$  &  $0.000260$ 		   &  $1.01$             \\
Rh	  &  $0.00848$         &  $0.0000987$  		  &  $0.00290$  &  $0.000293$ 		   &  $1.93$             \\
Sc	  &  $0.0274\,\,\,$    &  $0.000159\,\,\,$	  &  $0.00500$  &  $0.000229$ 		   &  $0.21$             \\
Sr	  &  $0.0205\,\,\,$    &  $0.000117\,\,\,$	  &  $0.00577$  &  $0.000274$ 		   &  $0.91$             \\
Ta	  &  $0.0110\,\,\,$    &  $0.000100\,\,\,$    &  $0.00274$  &  $0.000258$ 		   &  $1.92$             \\
Ti	  &  $0.0199\,\,\,$    &  $0.000121\,\,\,$	  &  $0.00456$  &  $0.000272$ 		   &  $0.49$             \\
Tm	  &  $0.0192\,\,\,$    &  $0.000140\,\,\,$	  &  $0.00354$  &  $0.000212$ 		   &  $0.58$             \\
V	  &  $0.0185\,\,\,$    &  $0.000140\,\,\,$	  &  $0.00329$  &  $0.000222$ 		   &  $0.46$             \\
W	  &  $0.0153\,\,\,$    &  $0.000124\,\,\,$	  &  $0.00221$  &  $0.000206$ 		   &  $1.01$             \\
Zr	  &  $0.0177\,\,\,$    &  $0.000120\,\,\,$	  &  $0.00389$  &  $0.000273$ 		   &  $0.66$             \\ \hline\hline
\end{tabular}
\end{table}

The semi-empirical expression remains highly accurate within $[0,1000]\,$nm regardless of the metal, but the mean absolute relative errors visibly increase for each element, now ranging from $0.21\%$ (Sc) up to $3.24\%$ (Al) with an element-averaged accuracy of $1.27\%$. It is worth pointing out that the semi-empirical expression becomes rather inaccurate when the separation grid is further expanded up to $l\in[0,5]\,\mu$m, exhibiting an element-averaged accuracy of $8.71\%$. This stems from the increasing importance of the static magneto-dielectric contribution with increasing separation, which is not incorporated in the semi-empirical expression. In order words, the separations are large enough to be located at the crossover of the dispersion force regime from the fully retarded limit to the thermal limit. In fact, the largest relative errors in the $l\in[0,5]\,\mu$m interval are observed for the ferromagnetic metals (Fe, Ni, Co), since these elements possess the largest static contribution due to the enhanced static component of magnetic origin. The modification of the semi-empirical expression to
\begin{align*}
A(l)=\frac{1}{a+bl}+\frac{1}{c+dl}+\frac{3}{4}k_{\mathrm{b}}T\zeta(3)+\frac{3}{4}k_{\mathrm{b}}T\mathrm{Li}_3\left(\Delta_{13}^{\mathrm{s}}\Delta_{23}^{\mathrm{s}}\right)\,,
\end{align*}
improves the rather poor accuracy at extended separations but deteriorates the excellent accuracy at short separations, thus it will not be discussed further. 

\begin{figure*}[!t]
	\centering
	\includegraphics[width=6.45in]{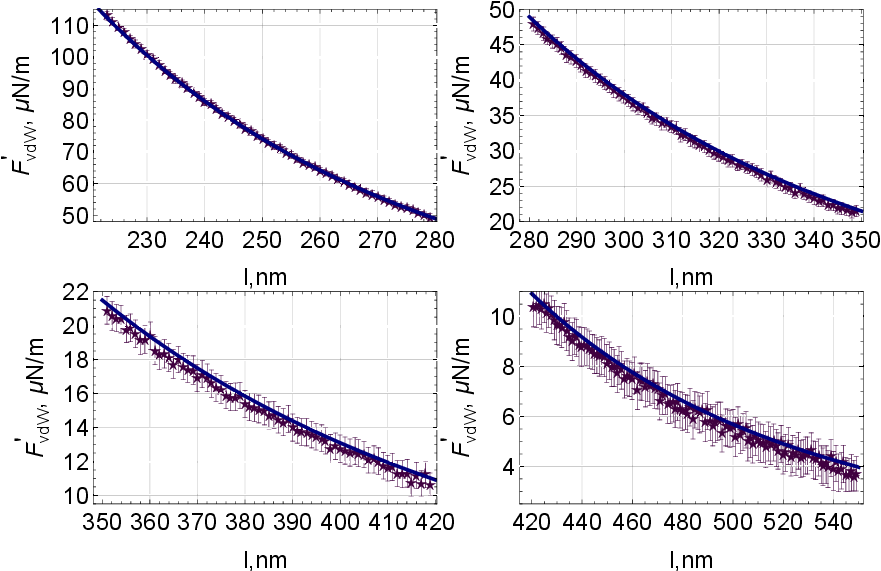}
	\caption{Gradient of the dispersion force between a $61.71\,\mu$m radius Ni sphere and a Ni plate as measured with dynamic atomic force microscopy\,\cite{measure7,measure8} (purple stars and purple error bars) and as computed from our semi-empirical formula within the proximity force approximation (blue solid line), see Eq.(\ref{CasimirNickel}) with input from Table \ref{hamakermicronvacuum}, at the separation range $l\in[220,550]\,$nm.}\label{fig:CasimirMeasurement}
\end{figure*}

This extension allows us to compare the predictions of our semi-empirical expression to high precision dispersion force measurements at submicrometer separation distances \cite{measure1,measure2,measure3,measure4,measure5,measure6,measure7,measure8,measure9}. We shall focus on dynamic atomic force microscope measurements of the gradient of the dispersion force between a Ni-coated hollow glass microsphere of $R_{\mathrm{s}}=61.71\pm0.09\,\mu$m radius and a Ni-coated Si plate at the separation range $l\in[220,550]\,$nm\,\cite{measure7,measure8}. The thickness of the Ni film
is $210\pm1\,$nm and $250\pm1\,$nm for the sphere and the plate, which is thick enough so that both objects can be assumed to have a pure Ni composition. In addition, the ratio $l/R_{\mathrm{s}}$ is sufficiently small so that the Derjaguin approximation (often referred to also as the proximity force approximation) is applicable\,\cite{proximi1,proximi2}. For the perfect sphere - semispace configuration, the Derjaguin approximation simply yields the force expression $F_{\mathrm{vdW}}=2\pi{R}_{\mathrm{s}}g(l)$, where $g(l)$ is the interaction free energy per unit area for the semispace - semispace configuration. Since the measurements of interest measure the derivative of the dispersion force, substitution of our semi-empirical expression to the above, leads to
\begin{align}
F^{\prime}_{\mathrm{vdW}}(l)&=\frac{R_{\mathrm{s}}}{6}\left[\frac{2al+3bl^2}{(al^2+bl^3)^2}+\frac{2cl+3dl^2}{(cl^2+dl^3)^2}\right]\,,\label{CasimirNickel}
\end{align}
with $R_{\mathrm{s}}$ in nm, $l$ in nm and $F^{\prime}_{\mathrm{vdW}}$ in $\mu${N}/m units. The comparison with the experiment is featured in Fig.\ref{fig:CasimirMeasurement}. The theoretical results lie well within the experimental error bars at all the probed separations $l\in[220,550]\,$nm. Thus, in spite of the first derivative operation, our unconstrained semi-empirical formula leads to a truly exceptional agreement with the measurements. It is important to point out that the agreement was achieved within a Drude model extrapolation, which contradicts the conclusion of Refs.\cite{measure7,measure8} that a plasma model extrapolation is required for an agreement with the force gradient measurements. Given the fact that the same extended-in-frequency optical data are essentially employed in both works, the difference lies in the number of data points that were selected for the low frequency extrapolation, which leads to different plasma frequencies and damping constants. The Drude model extrapolation versus plasma model extrapolation lies at the heart of an ongoing scientific debate\,\cite{Buhmann3,debateC1,debateC2,debateC3} that we intend to revisit in the future, but lies beyond the scope of the present work.

\begin{table}[!t]
\centering
\caption{Parameterizations of the room temperature Hamaker coefficients between $26$ identical elemental polycrystalline metals that are embedded in water, valid for separations within $0-200\,$nm, as computed from full Lifshitz theory [see Eqs.(\ref{Hamakergeneral1},\ref{Hamakergeneral2},\ref{Hamakergeneral3},\ref{Hamakergeneral4},\ref{Hamakergeneral5},\ref{Hamakergeneral6})]. The four coefficients $a,\,b,\,c,\,d$ of the unconstrained Eq.(\ref{hamakerfit}) are determined by least square fitting for separations within $0-200\,$nm. The mean absolute relative errors of the fits are reported for each element. The Fiedler \emph{et al.} parameterization is followed for the dielectric representation of water\cite{waterdi5}.}\label{hamakernanoFiedler}
\begin{tabular}{c c c c c c c c}
\\ \hline\hline
(F)   &	       a             &     b	                    &    c            &     d		         &  $e_{\mathrm{r}}$   \\
	  &  (zJ)$^{-1}$         & (zJ$\cdot$nm)$^{-1}$         & (zJ)$^{-1}$     & (zJ$\cdot$nm)$^{-1}$ &  $\%$               \\ \hline\hline
Ag    &  $0.0129\,\,\,$      &  $0.000109\,\,\,$  	     	&  $0.00731$      &  $0.000732$  		 &  $0.25$             \\	
Al	  &  $0.00749$           &  $0.000121\,\,\,$  		    &  $0.00907$      &  $0.000421$  		 &  $0.40$             \\	
Au	  &  $0.0170\,\,\,$      &  $0.000120\,\,\,$  	      	&  $0.00549$      &  $0.000481$  		 &  $0.25$             \\	
Ba	  &  $0.347\,\,\,\,\,\,$ &  $0\qquad\quad\,\,\,\,\,$    &  $0.0109\,\,\,$ &  $0.000168$  		 &  $0.28$             \\
Be	  &  $0.0494\,\,\,$      &  $0.000161\,\,\,$    		&  $0.00476$      &  $0.000189$  		 &  $0.43$             \\		
Co	  &  $0.0264\,\,\,$      &  $0.000130\,\,\,$     		&  $0.00459$      &  $0.000314$  		 &  $0.09$             \\			
Cr 	  &  $0.0258\,\,\,$      &  $0.000165\,\,\,$    		&  $0.00526$      &  $0.000274$  		 &  $0.23$             \\
Cu	  &  $0.0185\,\,\,$      &  $0.000124\,\,\,$  	      	&  $0.00720$      &  $0.000465$  		 &  $0.18$             \\					
Fe	  &  $0.0248\,\,\,$      &  $0.000117\,\,\,$  	      	&  $0.00498$      &  $0.000302$  		 &  $0.12$             \\
Hf	  &  $0.0231\,\,\,$      &  $0.000253\,\,\,$    		&  $0.0110\,\,\,$ &  $0.000653$  		 &  $0.22$             \\	
Ir	  &  $0.0134\,\,\,$      &  $0.000147\,\,\,$    		&  $0.00341$      &  $0.000326$  		 &  $0.32$             \\
Mo	  &  $0.0192\,\,\,$      &  $0.000144\,\,\,$    		&  $0.00366$      &  $0.000269$  		 &  $0.39$             \\	
Nb	  &  $0.0223\,\,\,$      &  $0.000141\,\,\,$    		&  $0.00387$      &  $0.000295$  		 &  $0.38$             \\	
Ni	  &  $0.0214\,\,\,$      &  $0.000127\,\,\,$     		&  $0.00575$      &  $0.000375$  		 &  $0.14$             \\	
Os	  &  $0.0383\,\,\,$      &  $0.000250\,\,\,$    		&  $0.00356$      &  $0.000295$  		 &  $0.11$             \\	
Pd	  &  $0.0192\,\,\,$      &  $0.000181\,\,\,$    		&  $0.00566$      &  $0.000459$  		 &  $0.08$             \\
Pt	  &  $0.0188\,\,\,$      &  $0.000162\,\,\,$   	     	&  $0.00432$      &  $0.000371$  		 &  $0.12$             \\
Rh	  &  $0.0136\,\,\,$      &  $0.000141\,\,\,$  	      	&  $0.00446$      &  $0.000373$  		 &  $0.28$             \\
Sc	  &  $0.257\,\,\,\,\,\,$ &  $0\qquad\quad\,\,\,\,\,$    &  $0.00846$      &  $0.000207$  		 &  $0.29$             \\
Sr	  &  $0.113\,\,\,\,\,\,$ &  $0.0000989$   		        &  $0.00949$      &  $0.000209$  		 &  $0.35$             \\
Ta	  &  $0.0185\,\,\,$      &  $0.000132\,\,\,$  	      	&  $0.00425$      &  $0.000338$  		 &  $0.29$             \\
Ti	  &  $0.0513\,\,\,$      &  $0.000178\,\,\,$    		&  $0.00828$      &  $0.000285$  		 &  $0.16$             \\
Tm	  &  $0.0690\,\,\,$      &  $0.000194\,\,\,$    		&  $0.00552$      &  $0.000225$  		 &  $0.25$             \\
V	  &  $0.0469\,\,\,$      &  $0.000214\,\,\,$    		&  $0.00532$      &  $0.000260$  		 &  $0.16$             \\
W	  &  $0.0295\,\,\,$      &  $0.000159\,\,\,$    		&  $0.00325$      &  $0.000275$  		 &  $0.22$             \\
Zr	  &  $0.0305\,\,\,$      &  $0.000176\,\,\,$    		&  $0.00729$      &  $0.000357$  		 &  $0.15$             \\ \hline\hline
\end{tabular}
\end{table}

\begin{table}[!t]
\centering
\caption{Parameterizations of the room temperature Hamaker coefficients between $26$ identical elemental polycrystalline metals that are embedded in water, valid for separations within $0-200\,$nm, as computed from full Lifshitz theory [see Eqs.(\ref{Hamakergeneral1},\ref{Hamakergeneral2},\ref{Hamakergeneral3},\ref{Hamakergeneral4},\ref{Hamakergeneral5},\ref{Hamakergeneral6})]. The four coefficients $a,\,b,\,c,\,d$ of the unconstrained Eq.(\ref{hamakerfit}) are determined by least square fitting for separations within $0-200\,$nm. The mean absolute relative errors of the fits are reported for each element. The Parsegian-Weiss parameterization is followed for the dielectric representation of water\cite{waterdi1}.}\label{hamakernanoPW}
\begin{tabular}{c c c c c c c c}
\\ \hline\hline
(PW)  &	       a             &     b	                  &    c             &     d		          &   $e_{\mathrm{r}}$  \\
	  &  (zJ)$^{-1}$         & (zJ$\cdot$nm)$^{-1}$       & (zJ)$^{-1}$      & (zJ$\cdot$nm)$^{-1}$   &  $\%$              \\ \hline\hline
Ag    &  $0.0128\,\,\,$      &  $0.000107\,\,\,$  		  &  $0.00609$       &  $0.000691$  		  &  $0.29$             \\	
Al	  &  $0.00700$           &  $0.000114\,\,\,$  		  &  $0.00859$       &  $0.000468$  		  &  $0.41$             \\	
Au	  &  $0.0165\,\,\,$      &  $0.000117\,\,\,$  		  &  $0.00475$       &  $0.000472$  		  &  $0.27$             \\	
Ba	  &  $0.246\,\,\,\,\,\,$ &  $0\qquad\quad\,\,\,\,\,$  &  $0.0103\,\,\,$  &  $0.000175$  		  &  $0.26$             \\
Be	  &  $0.0374\,\,\,$      &  $0.000161\,\,\,$    	  &  $0.00451$       &  $0.000201$  		  &  $0.39$             \\		
Co	  &  $0.0234\,\,\,$      &  $0.000129\,\,\,$  		  &  $0.00413$       &  $0.000326$  		  &  $0.12$             \\			
Cr 	  &  $0.0203\,\,\,$      &  $0.000155\,\,\,$    	  &  $0.00492$       &  $0.000308$  		  &  $0.17$             \\
Cu	  &  $0.0168\,\,\,$      &  $0.000120\,\,\,$  		  &  $0.00635$       &  $0.000499$  		  &  $0.14$             \\				
Fe	  &  $0.0215\,\,\,$      &  $0.000116\,\,\,$  		  &  $0.00452$       &  $0.000322$  		  &  $0.09$             \\
Hf	  &  $0.0185\,\,\,$      &  $0.000231\,\,\,$    	  &  $0.00979$       &  $0.000812$  		  &  $0.25$             \\	
Ir	  &  $0.0131\,\,\,$      &  $0.000143\,\,\,$    	  &  $0.00306$       &  $0.000319$  		  &  $0.33$             \\
Mo	  &  $0.0177\,\,\,$      &  $0.000139\,\,\,$    	  &  $0.00335$       &  $0.000273$  		  &  $0.38$             \\	
Nb	  &  $0.0208\,\,\,$      &  $0.000137\,\,\,$    	  &  $0.00349$       &  $0.000296$  		  &  $0.35$             \\	
Ni	  &  $0.0186\,\,\,$      &  $0.000124\,\,\,$  		  &  $0.00517$       &  $0.000407$  		  &  $0.20$             \\	
Os	  &  $0.0338\,\,\,$      &  $0.000244\,\,\,$    	  &  $0.00321$       &  $0.000296$  		  &  $0.13$             \\	
Pd	  &  $0.0179\,\,\,$      &  $0.000175\,\,\,$    	  &  $0.00495$       &  $0.000465$  		  &  $0.08$             \\
Pt	  &  $0.0178\,\,\,$      &  $0.000156\,\,\,$    	  &  $0.00384$       &  $0.000371$  		  &  $0.12$             \\
Rh	  &  $0.0129\,\,\,$      &  $0.000136\,\,\,$  		  &  $0.00398$       &  $0.000375$  		  &  $0.30$             \\
Sc	  &  $0.00804$           &  $0.000240\,\,\,$	   	  &  $0.0956\,\,\,$  &  $0.000247$  		  &  $0.17$             \\
Sr	  &  $0.0521\,\,\,$      &  $0.0001659$   		      &  $0.00942$       &  $0.000266$  		  &  $0.23$             \\
Ta	  &  $0.0174\,\,\,$      &  $0.000128\,\,\,$  		  &  $0.00381$       &  $0.000340$  		  &  $0.28$             \\
Ti	  &  $0.0292\,\,\,$      &  $0.000175\,\,\,$    	  &  $0.00810$       &  $0.000381$  		  &  $0.07$             \\
Tm	  &  $0.0405\,\,\,$      &  $0.000208\,\,\,$    	  &  $0.00522$       &  $0.000257$  		  &  $0.14$             \\
V	  &  $0.0315\,\,\,$      &  $0.000207\,\,\,$    	  &  $0.00497$       &  $0.000296$  		  &  $0.08$             \\
W     &  $0.0275\,\,\,$      &  $0.000156\,\,\,$    	  &  $0.00295$       &  $0.000273$  		  &  $0.20$             \\
Zr	  &  $0.0236\,\,\,$      &  $0.000167\,\,\,$    	  &  $0.00671$       &  $0.000417$  		  &  $0.09$             \\ \hline\hline
\end{tabular}
\end{table}

\subsection{Identical metals in pure water}\label{numericalwater}

\noindent The retarded room temperature Hamaker coefficients of the van der Waals interactions between identical isotropic elemental metals embedded in pure water have been computed from the full Lifshitz theory for the three dielectric representations of water. The Lifshitz calculations focused on semi-space separations $l\in[0,200]\,$nm and the exact results were fitted to the unconstrained semi-empirical Eq.(\ref{hamakerfit}). The least-square fit parameters $a,b,c,d$ and the accuracy of the simple unconstrained semi-empirical expression are reported in Tables \ref{hamakernanoFiedler}, \ref{hamakernanoPW}, \ref{hamakernanoRL} for the Fiedler \emph{et al.}, the Parsegian-Weiss and the Roth-Lenhoff representations, respectively. Regardless of the dielectric parameterization of choice, it is apparent that the semi-empirical expression is remarkably accurate for separations $l\in[0,200]$ nm. For the Fiedler \emph{et al.} representation, the mean absolute relative errors of Eq.(\ref{hamakerfit}) range from $0.08\%$ (Pd) up to $0.43\%$ (Be) and the element-averaged accuracy is $0.24\%$. For the Parsegian-Weiss representation, the mean absolute relative errors of Eq.(\ref{hamakerfit}) range from $0.07\%$ (Ti) up to $0.41\%$ (Al) and the element-averaged accuracy is $0.21\%$. For the Roth-Lenhoff parameterization, the mean absolute relative errors of Eq.(\ref{hamakerfit}) range from $0.04\%$ (Ti) up to $0.42\%$ (Al) and the element-averaged accuracy is $0.22\%$. Considering also the results for vacuum, it appears that the unconstrained semi-empirical expression is very accurate at short and intermediate separations regardless of the elemental metal and intervening medium.

\begin{figure}[!t]
	\centering
	\includegraphics[width=3.40in]{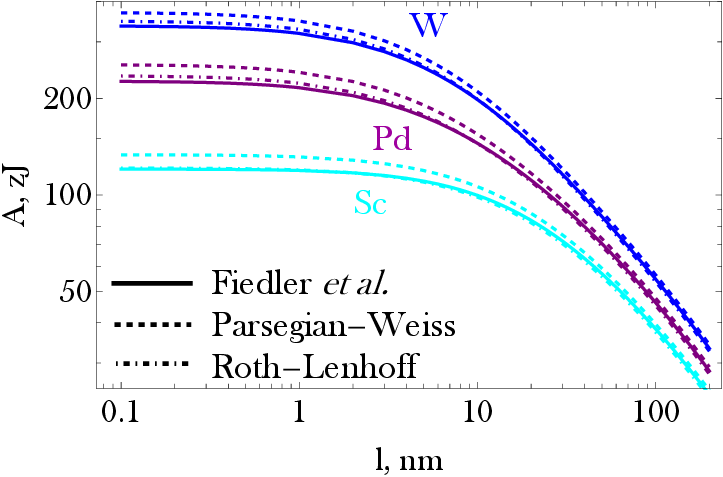}
	\caption{The room temperature Hamaker coefficients for W-W,\,Pd-Pd and Sc-Sc semi-spaces embedded in pure water, computed from the full Lifshitz theory with dielectric function input from the Fiedler \emph{et al.} representation, the Parsegian-Weiss representation and the Roth-Lenhoff representation, as functions of the separation.}\label{fig:metal_water}
\end{figure}

Regardless of elemental metal composition and separation, the Hamaker coefficients computed with input from the Parsegian-Weiss representation have the largest values, while the Hamaker coefficients computed with input from the Fiedler \emph{et al.} representation have the smallest values. In addition, the Hamaker coefficients computed with input from the Fiedler \emph{et al.} and Roth-Lenhoff representations have very similar values. Fig.\ref{fig:metal_water} contains some characteristic examples for identical metal combinations that belong to different columns of the periodic table.

\begin{table}[!t]
\centering
\caption{Parameterizations of the room temperature Hamaker coefficients between $26$ identical elemental polycrystalline metals that are embedded in water, valid for separations within $0-200\,$nm, as computed from full Lifshitz theory [see Eqs.(\ref{Hamakergeneral1},\ref{Hamakergeneral2},\ref{Hamakergeneral3},\ref{Hamakergeneral4},\ref{Hamakergeneral5},\ref{Hamakergeneral6})]. The four coefficients $a,\,b,\,c,\,d$ of the unconstrained Eq.(\ref{hamakerfit}) are determined by least square fitting for separations within $0-200\,$nm. The mean absolute relative errors of the fits are reported for each element. The Roth-Lenhoff parameterization is followed for the dielectric representation of water\cite{waterdi2}.}\label{hamakernanoRL}
\begin{tabular}{c c c c c c c c}
\\ \hline\hline
(RL)  &	       a             &     b	                   &    c              &     d		          &  $e_{\mathrm{r}}$   \\
	  &  (zJ)$^{-1}$         & (zJ$\cdot$nm)$^{-1}$        & (zJ)$^{-1}$       & (zJ$\cdot$nm)$^{-1}$ &  $\%$               \\ \hline\hline
Ag    &  $0.0129\,\,\,$      &  $0.000111\,\,\,$  		   &  $0.00685$        &  $0.000779$  		  &  $0.27$             \\	
Al	  &  $0.00738$           &  $0.000121\,\,\,$  		   &  $0.00906$        &  $0.000460$  		  &  $0.42$             \\	
Au	  &  $0.0167\,\,\,$      &  $0.000122\,\,\,$  	       &  $0.00523$        &  $0.000515$  		  &  $0.25$             \\	
Ba	  &  $0.345\,\,\,\,\,\,$ &  $0\qquad\quad\,\,\,\,\,$   &  $0.0111\,\,\,$   &  $0.000173$  		  &  $0.25$             \\
Be	  &  $0.0436\,\,\,$      &  $0.000168\,\,\,$    	   &  $0.00475$        &  $0.000202$  		  &  $0.41$             \\		
Co	  &  $0.0245\,\,\,$      &  $0.000133\,\,\,$  		   &  $0.00447$        &  $0.000343$  		  &  $0.12$             \\			
Cr 	  &  $0.0222\,\,\,$      &  $0.000164\,\,\,$    	   &  $0.00526$        &  $0.000312$  		  &  $0.18$             \\
Cu	  &  $0.0171\,\,\,$      &  $0.000125\,\,\,$  		   &  $0.00708$        &  $0.000539$  		  &  $0.15$             \\				
Fe	  &  $0.0227\,\,\,$      &  $0.000120\,\,\,$  		   &  $0.00488$        &  $0.000334$  		  &  $0.08$             \\
Hf	  &  $0.0188\,\,\,$      &  $0.000240\,\,\,$           &  $0.0115\,\,\,$   &  $0.000929$  		  &  $0.29$             \\	
Ir	  &  $0.0136\,\,\,$      &  $0.000150\,\,\,$    	   &  $0.00325$        &  $0.000335$  		  &  $0.33$             \\
Mo	  &  $0.0187\,\,\,$      &  $0.000146\,\,\,$    	   &  $0.00355$        &  $0.000283$  		  &  $0.38$             \\	
Nb	  &  $0.0217\,\,\,$      &  $0.000143\,\,\,$    	   &  $0.00374$        &  $0.000311$  		  &  $0.36$             \\	
Ni	  &  $0.0194\,\,\,$      &  $0.000129\,\,\,$  		   &  $0.00565$        &  $0.000428$  		  &  $0.22$             \\	
Os	  &  $0.0362\,\,\,$      &  $0.000254\,\,\,$    	   &  $0.00343$        &  $0.000312$  		  &  $0.13$             \\	
Pd	  &  $0.0182\,\,\,$      &  $0.000183\,\,\,$    	   &  $0.00546$        &  $0.000505$  		  &  $0.09$             \\
Pt	  &  $0.0183\,\,\,$      &  $0.000164\,\,\,$   	       &  $0.00415$        &  $0.000395$  		  &  $0.11$             \\
Rh	  &  $0.0133\,\,\,$      &  $0.000143\,\,\,$  		   &  $0.00429$        &  $0.000395$  		  &  $0.28$             \\
Sc	  &  $0.172\,\,\,\,\,\,$ &  $0.000152\,\,\,$	   	   &  $0.00856$        &  $0.000227$  		  &  $0.22$             \\
Sr	  &  $0.0748\,\,\,$      &  $0.0001649$   		       &  $0.00991$        &  $0.000243$  		  &  $0.27$             \\
Ta	  &  $0.0180\,\,\,$      &  $0.000133\,\,\,$  		   &  $0.00410$        &  $0.000360$  		  &  $0.28$             \\
Ti	  &  $0.0341\,\,\,$      &  $0.000185\,\,\,$    	   &  $0.00877$        &  $0.000370$  		  &  $0.04$             \\
Tm	  &  $0.0493\,\,\,$      &  $0.000217\,\,\,$    	   &  $0.00557$        &  $0.000256$  		  &  $0.15$             \\
V 	  &  $0.0358\,\,\,$      &  $0.000218\,\,\,$           &  $0.00534$        &  $0.000300$  		  &  $0.09$             \\
W     &  $0.0290\,\,\,$      &  $0.000162\,\,\,$    	   &  $0.00314$        &  $0.000287$  		  &  $0.21$             \\
Zr	  &  $0.0250\,\,\,$      &  $0.000176\,\,\,$           &  $0.00741$        &  $0.000433$  		  &  $0.07$             \\ \hline\hline
\end{tabular}
\end{table}

\begin{figure}
	\centering
	\includegraphics[width=3.40in]{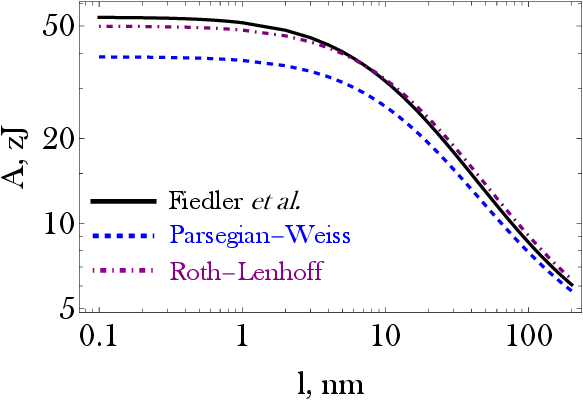}
	\caption{The room temperature Hamaker coefficients for pure water-water semi-spaces embedded in vacuum, computed from the full Lifshitz theory with dielectric function input from the Fiedler \emph{et al.} representation, the Parsegian-Weiss representation and the Roth-Lenhoff representation, as functions of the separation.}\label{fig:pure_water}
\end{figure}

These observations can be explained by the functional dependence of the auxiliary functions $\bar{\Delta}_{ij}(\cdot),\,x_i(\cdot)$ on the imaginary argument dielectric function of the medium, the functional dependence of the lower integration limit function $r_n(\cdot)$ on the imaginary argument dielectric function of the medium and the Hamaker coefficients of the water-vacuum-water system for the three dielectric representations. These Hamaker coefficients are illustrated in Fig.\ref{fig:pure_water}, where it becomes clear that $A_{\mathrm{wvw}}^{\mathrm{F}}(l)\gtrsim{A}_{\mathrm{wvw}}^{\mathrm{RL}}(l)>{A}_{\mathrm{wvw}}^{\mathrm{PW}}(l)$ holds for any semi-space separation. Courtesy of the functional dependence of $\bar{\Delta}_{ij}(\cdot),\,x_i(\cdot),\,r_n(\cdot)$ on $\epsilon(\imath\xi_n)$, this inequality should be inverted during ${A}_{\mathrm{1w2}}(l)$ full Lifshitz calculations, as indeed observed in Fig.\ref{fig:metal_water}.

Finally, it is worth noting that the Hamaker coefficients $A_{\mathrm{wvw}}^{\mathrm{F}}(l),\,{A}_{\mathrm{wvw}}^{\mathrm{RL}}(l),\,{A}_{\mathrm{wvw}}^{\mathrm{PW}}(l)$ strongly conform to the semi-empirical Eq.(\ref{hamakerfit}) with least-square fit parameters. Nevertheless, the semi-empirical expression for pure water has an accuracy level of $\gtrsim2\%$, which is rather substantially lower than its accuracy level for any elemental metal. This is a direct consequence of the relatively large value of the ratio of the static contribution to the Hamaker coefficient over the total Hamaker coefficient for water. As a result, the contribution of the static contribution is important even for small-to-intermediate separations and should be incorporated in the semi-empirical expression. Simultaneously, given the rather restricted-in-frequency ($0-25\,$eV to $0-100\,$eV) dielectric data for pure water compared to the extended-in-frequency ($0-10000\,$eV) for elemental metals, a single simple fraction can be employed. The above lead to the semi-empirical expression
\begin{align}
A_{\mathrm{wvw}}(l)=a+\frac{1}{b+cl}\,.\label{hamakerfitwater}
\end{align}
with the least-square parameters $a=3.129\,$zJ, $b=0.01921$ (zJ)$^{-1}$, $c=0.001634\,$(zJ$\cdot$nm)$^{-1}$ for the Fiedler \emph{et al.} representation with a $0.59\%$ mean absolute relative error, the least-square parameters $a=3.179\,$zJ, $b=0.02712\,$(zJ)$^{-1}$ and $c=0.001835\,$(zJ$\cdot$nm)$^{-1}$ for the Parsegian-Weiss representation with a $0.65\%$ mean absolute relative error, the least-square parameters $a=3.097\,$zJ, $b=0.02066\,$(zJ)$^{-1}$, $c=0.001466\,$(zJ$\cdot$nm)$^{-1}$ for the Roth-Lenhoff representation with a $0.63\%$ mean absolute relative error.

\section{Summary and importance}\label{outro}

\noindent Exact Lifshitz calculations have been reported for the retarded room temperature Hamaker coefficients between $26$ identical isotropic polycrystalline metals embedded in vacuum and pure water for semi-space separations primarily within $l\in[0,200]\,$nm but also within $l\in[0,1000]\,$nm. The computation of the imaginary argument dielectric function of metals was based on the full spectral method combined with a Drude model low frequency extrapolation technique which has been implemented with input from extended-in-frequency dielectric data that range from the far infra-red region to the soft X-ray region of the electromagnetic spectrum. The computation of the imaginary argument dielectric function of pure water was based on the simple spectral method which has been implemented with input from three different dielectric parameterizations; the Fiedler \emph{et al.} representation, the Parsegian-Weiss representation and the Roth-Lenhoff representation.

In addition, the retarded room temperature Hamaker coefficients between $26$ identical isotropic polycrystalline metals that are embedded in vacuum have also been computed with five common approximations of the Lifshitz theory, \emph{i.e.}, the low temperature approximation, the modified low temperature approximation, the dipole approximation, the low temperature dipole approximation and the modified low temperature dipole approximation, for semi-space separations within $l\in[0,100]\,$nm. Moreover, a compact yet very accurate semi-empirical expression has been proposed to describe the separation dependence of the retarded room temperature Hamaker coefficient between any metals embedded in vacuum or pure water. The semi-empirical expression has been constructed by incorporating the well-known exact asymptotic limits of the Hamaker coefficient in the simplest manner and features only four least-square fitted parameters. Finally, the utility of the semi-empirical expression has been demonstrated through a comparison with high precision dispersion force measurements between Ni spheres and Ni plates, where an excellent agreement was revealed.

The primary contributions of this investigation are the following. \emph{First and foremost}, the present extensive compilation features the most accurate retarded room temperature Hamaker coefficients ever reported in the literature. Therefore, the computed Hamaker coefficients can be compared with dedicated van der Waals measurements\,\cite{outrome1,outrome2,outrome3}, adopted directly in the modelling of the van der Waals interactions in different physical phenomena\,\cite{outromo1,outromo2,outromo3}, employed for the benchmarking of dedicated software implementation of Lifshitz theory (see for instance the Gecko Hamaker tool)\,\cite{outroGe1,outroGe2,outroGe3} but also utilized as reference values in advanced theoretical studies of implicit temperature effects in bulk materials\,\cite{outroNa1}, size effects in nano-particles\,\cite{outroNa1,outroNa2}, spatial dispersion effects in bulk materials\,\cite{outroSp1,outroSp2}, beyond step-like interface effects\,\cite{outroSt1,outroSt2} and inhomogeneity effects\,\cite{outroIn1,outroIn2,outroIn3}. \emph{Furthermore}, the modified low temperature approximation, at least for semi-space separations within $l\in[0,100]\,$nm, has proven to be extremely accurate for all metals of interest with mean absolute relative deviations $0.53\%$ (element-averaged) from full Lifshitz theory. Given the fact that the dielectric input cannot be expected to be accurate within $1\%$, this implies that the modified low temperature approximation can be employed instead of the full Lifshitz theory, which leads to a substantial reduction of the computational cost. \emph{Finally}, the simple novel unconstrained semi-empirical expression, at least for semi-space separations within $l\in[0,200]\,$nm, has proven to be extremely accurate with mean absolute relative deviations $<0.5\%$ (for all elemental metals and intervening media of interest) from the Hamaker coefficients that result from full Lifshitz theory. Again, given the fact that the dielectric input cannot be expected to be accurate within $1\%$, the semi-empirical expression can reliably substitute the cumbersome calculations of the full Lifshitz theory. It also constitutes an extremely effective, compression-wise, representation of Hamaker coefficients that can substitute lengthy tabulations.

Future work will mainly focus on two directions. First, the highly accurate semi-empirical Hamaker coefficient description will be expanded at separations beyond the micrometer range; a task which requires an analytical expression that is able to capture the cross-over behavior between the non-retarded, fully retarded and thermal regimes\,\cite{outrocro}. Second, at small distances of the order of few nanometers, electromagnetic modes of wavelength that is comparable to the interatomic spacing provide important contributions, which implies that the long wavelength magneto-dielectric response functions do not suffice and that spatial dispersion also needs to be considered\,\cite{outroSp2}. It is evident that the incorporation of spatial dispersion effects either requires a finite-wavenumber extension of the experimental optical data\,\cite{outroPen} or the adaptation of a first-principle dielectric metal response model.

\end{document}